\documentclass{WileyMSP-template}

\usepackage{graphicx,a4wide,url}
\usepackage{hyperref}
\usepackage{xcolor}
\usepackage{amsmath}
\usepackage{amssymb}
\usepackage{amsthm}
\usepackage{enumerate}
\usepackage{floatrow}
\usepackage{subfig}
\usepackage{textgreek}
\usepackage{algorithm}
\usepackage{algpseudocode}
\usepackage{lineno}
\usepackage{listings}
\usepackage{longtable}
\usepackage{parskip}
\usepackage{pythonhighlight}

\graphicspath{ {FIGURES/} }

\lstdefinestyle{terminal}{
language=python,
showtabs=true,
tab=,
tabsize=2,
basicstyle=\ttfamily\footnotesize,
stringstyle=\color{stringcolour},
showstringspaces=false,
alsoletter={1234567890},
otherkeywords={\%, \}, \{, \&, \|},
keywordstyle=\color{keywordcolour}\bfseries,
emph={and,break,class,continue,def,yield,del,elif ,else,%
except,exec,finally,for,from,global,if,import,in,%
lambda,not,or,pass,print,raise,return,try,while,assert,with},
emphstyle=\color{blue}\bfseries,
emph={[2]True, False, None},
emphstyle=[2]\color{keywordcolour},
emph={[3]object,type,isinstance,copy,deepcopy,zip,enumerate,reversed,list,set,len,dict,tuple,xrange,append,execfile,real,imag,reduce,str,repr},
emphstyle=[3]\color{commandcolour},
emph={Exception,NameError,IndexError,SyntaxError,TypeError,ValueError,OverflowError,ZeroDivisionError},
emphstyle=\color{exceptioncolour}\bfseries,
morecomment=[s]{"""}{"""},
commentstyle=\color{commentcolour}\slshape,
emph={[4]ode, fsolve, sqrt, exp, sin, cos,arctan, arctan2, arccos, pi,  array, norm, solve, dot, arange, isscalar, max, sum, flatten, shape, reshape, find, any, all, abs, plot, linspace, legend, quad, polyval,polyfit, hstack, concatenate,vstack,column_stack,empty,zeros,ones,rand,vander,grid,pcolor,eig,eigs,eigvals,svd,qr,tan,det,logspace,roll,min,mean,cumsum,cumprod,diff,vectorize,lstsq,cla,eye,xlabel,ylabel,squeeze},
emphstyle=[4]\color{numpycolour},
emph={[5]__init__,__add__,__mul__,__div__,__sub__,__call__,__getitem__,__setitem__,__eq__,__ne__,__nonzero__,__rmul__,__radd__,__repr__,__str__,__get__,__truediv__,__pow__,__name__,__future__,__all__},
emphstyle=[5]\color{specmethodcolour},
emph={[6]assert,yield},
emphstyle=[6]\color{keywordcolour}\bfseries,
emph={[7]range},
emphstyle={[7]\color{keywordcolour}\bfseries},
literate=*%
{*}{{\literatecolour*}}{1}%
{**}{{\literatecolour{**}}}2%
{!}{{\literatecolour!}}{1}%
{[}{{\literatecolour[}}{1}%
{]}{{\literatecolour]}}{1}%
{<}{{\literatecolour<}}{1}%
{>}{{\literatecolour>}}{1}%
{>>>}{\pythonprompt}{3}%
,%
frame=trbl,
rulecolor=\color{black!40},
backgroundcolor=\color{white},
breakindent=.5\textwidth,frame=single,breaklines=true%
}

\lstnewenvironment{mypython}[1][]{\lstset{style=mypython,#1}}{}
\lstnewenvironment{terminal}[1][]{\lstset{style=terminal,#1}}{}
\newcommand{\revision}[1]{{\color{black}#1}}
\begin{document}

\pagestyle{fancy}

\title{Sphractal: Estimating the Fractal Dimension of Surfaces Computed from Precise Atomic Coordinates via Box-Counting Algorithm}

\maketitle

\author{Jonathan Yik Chang Ting,*}
\author{Andrew Thomas Agars Wood,}
\author{Amanda Susan Barnard}

\dedication{}

\begin{affiliations}
J. Y. C. Ting, Prof. A. S. Barnard \\
School of Computing, 145 Science Road, \\
Australian National University, 2601 Canberra, ACT, Australia.\\
Email Address: Jonathan.Ting@anu.edu.au \par

Prof. A. T. A. Wood \\
Research School of Finance, Actuarial Studies \& Statistics, 26C Kingsley Street, \\
Australian National University, 2601 Canberra, ACT, Australia.
\end{affiliations}


\begin{abstract}
The fractal dimension of a surface allows its degree of roughness to be characterised quantitatively. However, limited effort has been attempted to compute the fractal dimension of surfaces computed from precisely known atomic coordinates from computational biomolecular and nanomaterial studies. This work proposes methods to estimate the fractal dimension of the surface of any three-dimensional object composed of spheres, by representing it as either a voxelised point cloud or a mathematically exact surface, and computing its box-counting dimension. Sphractal is published as a Python package that provides these functionalities, and its utility is demonstrated on a set of simulated palladium nanoparticle data.
\end{abstract}

\section{Introduction}
\label{sec:intro}

\begin{figure}[htbp]  \centering
\sidesubfloat[]{\includegraphics[width=0.1\textwidth]{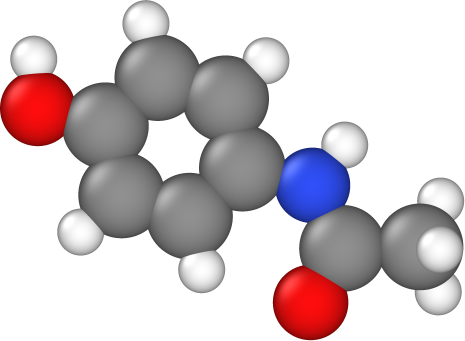}\label{fig:paracetamolSurface}}
\sidesubfloat[]{\includegraphics[width=0.15\textwidth]{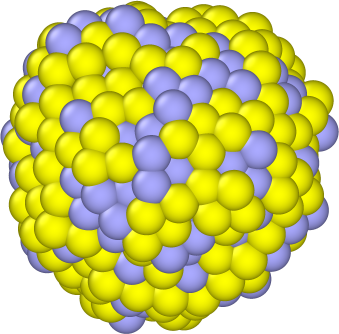}\label{fig:AuPdNPsurface}}
\sidesubfloat[]{\includegraphics[width=0.25\textwidth]{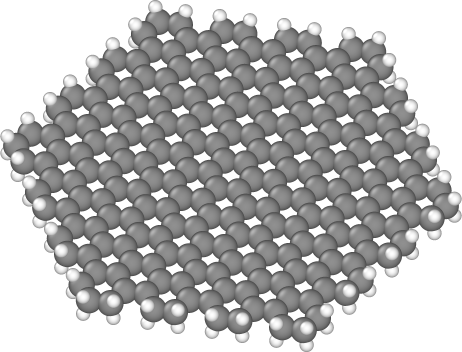}\label{fig:grapheneSurface}}
\sidesubfloat[]{\includegraphics[width=0.3\textwidth]{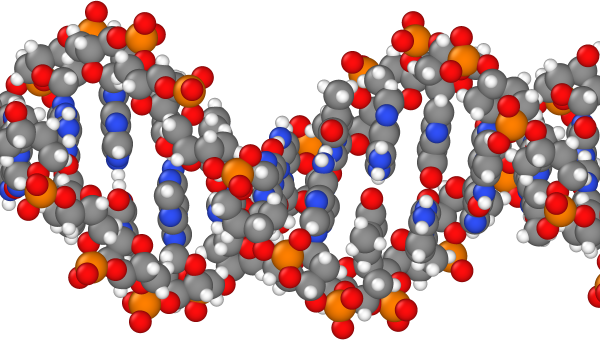}\label{fig:DNAsurface}} \\
\caption{Illustrations of the complex atomistic surfaces of a (a) paracetamol, (b) gold-palladium nanoparticle, (c) graphene nanosheet, and (d) deoxyribonucleic acid. The white, gray, blue, red, orange, yellow, and light blue spheres correspond to hydrogen, carbon, nitrogen, oxygen, phosphorus, gold, and palladium atoms, respectively.}
\label{fig:atomicSurfaces}
\end{figure}

In the computational studies of molecular and nanomaterial sciences, atomistic objects such as small molecules, proteins, nanoparticles, polymers, and porous materials are often modelled as collections of spheres sized according to their atomic or molecular radii. The resulting surfaces of these objects are depicted as complex shapes with intricate details due to overlapping spheres, as illustrated in Figure~\ref{fig:atomicSurfaces}. These details entail the protrusions (or convexities) and indentations (or concavities) on the surface and contribute to the overall surface roughness, which significantly affects the properties of the object~\cite{Cui2013surfCatalAct,Jeon2021surfORR}. This is because surface roughness determines the amount of surface area available for interaction with other entities (such as drug-protein and protein-protein interactions in drug design~\cite{Archakov2003PPI,Smith2012PPI}, protein-protein interaction in biomolecular mechanistic studies~\cite{Nooren2003PPI}, and reactant-catalyst interaction in catalysis~\cite{Somwanshi2020nanoCat,Yoo2022RCI}). However, there is a lack of tools to quantitatively describe such complex surfaces. \par

A solution is to leverage the extensive literature on the characterisation of object roughness or shape complexity~\cite{Backes2010model3DFD,Backes2010MSFD,Gardiner2018shapeComplexity}. Given the coordinates of the individual atoms in an atomistic object, the entire surface could be traced out by assuming that each atom is a sphere that overlaps with neighbouring atoms. This collective surface can be treated as a three-dimensional (3D) object, which can then be characterised by the surface roughness parameters from the shape complexity research domain. One such parameter is fractal dimension~\cite{Chan2004FD}. \par

\begin{figure}
  \centering
  \includegraphics[width=0.9\textwidth]{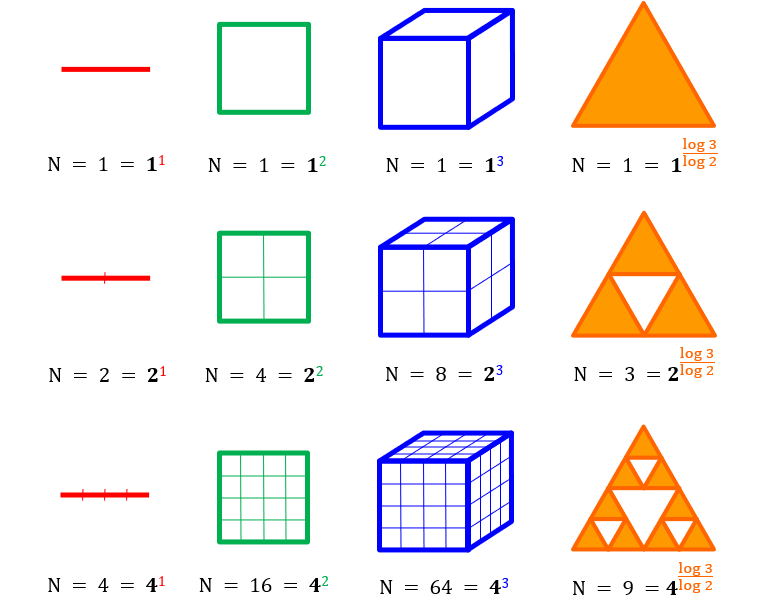}
  \caption{An illustration of the dimension concept based on a scaling relationship. N represents the number of measurement units, the bold number denotes the inverse of scaling factor, while the coloured exponent indicates the dimension of the object. The first 3 iterations of the Sierpi\'nski triangle are shown on the right-most column.}
  \label{fig:DimensionConcept}
\end{figure}

A fractal dimension is a unitless and scale-free real value index that quantifies the complexity of an object by measuring the ratio of detail change to scale change~\cite{Mandelbrot1982FDcoin}. This is derived from the understanding of the concept of dimension as a scaling relationship, mathematically defined as a general scaling rule (Equation~\ref{eq:Dimension}), as illustrated in Figure~\ref{fig:DimensionConcept}. Some examples of measurement units in Equation~\ref{eq:Dimension} include rulers used to measure the length of a 1D line, squares (pixels) or circles used to cover the area occupied by a 2D shape, and cubes (voxels) or spheres used to cover the volume occupied by a 3D object. \par

\begin{equation} \label{eq:Dimension}
N(\epsilon) = \epsilon^{-D} = \left( 1 / \epsilon \right) ^D
\end{equation}
where: 

\hspace{24pt} $N$ = number of measurement units 

\hspace{24pt} $\epsilon$ = scaling factor

\hspace{24pt} $D$ = dimension \par

While fractal dimension $D_{f}$ is a scale invariant measure (Equation~\ref{eq:FracDim}), in practice naturally occurring objects will only exhibit fractal-like behaviour over a finite range of scales. \revision{The objects of interest in this study are surfaces resulting from the union of a finite number of overlapping spheres. These objects have $D_{f}$ equal to $2$: zooming in at any smooth point on the surface, allowing $\epsilon$ to approach $0$, the limit is a plane, which has dimension $2$. Despite this, as shown in this paper, the methods used to calculate the fractal dimension of self-similar mathematical fractals are still useful when applied to such surfaces, as they provide us with a useful, heuristic, empirical measure of the roughness of the surfaces when applied in practice. To avoid cumbersome explanations in the following text, we will simply describe the action of applying the fractal dimension calculation on our objects of interest as estimating the fractal dimension of the objects.}

\begin{equation} \label{eq:FracDim}
D_{f} = \text{lim}_{\epsilon \rightarrow 0} \frac{\text{log}( N(\epsilon) )} {\text{log}( 1/\epsilon )}
\end{equation}

Researchers have used fractal dimension calculation methods to characterise galaxies~\cite{Yadav2010FDgalaxy}, landscapes~\cite{Hagerhall2004FDlandscape}, coastlines~\cite{Mandelbrot1967FDcoin}, clouds~\cite{Hentschel1984FDcloud}, lightnings~\cite{Sanudo1995FDlightning}, cerebellums~\cite{Marzi2018FDbrain}, cauliflowers~\cite{Zanoni2002FDcauliflower}, and leaves~\cite{Du2013FDleaf}. Numerous methods have been developed to estimate the fractal dimension of 3D objects, with the most common approach being the box-counting method~\cite{Russell1980boxCntFD} and its variations~\cite{Mandelbrot1982FDcoin,Sanderson1990yardstickFD,Chaudhuri1995diffBoxCntFD,Sandau1997extendCntFD}. This is due to (i) the ease of implementation~\cite{Yang2015vegetation3DFD,Wang2022aggregate3DFD,Wang2015sapphire3DFD} and high reproducibility~\cite{Goni2013boxCntReprod,Wang2022aggregate3DFD,Wang2015sapphire3DFD} of the method, (ii) applicability of the method for objects that are not exactly self-similar~\cite{Zhang2006boxCnt,Zhang2007boxCnt,Wu2010MRI3DFD,Pashminehazar2019agglomerate3DFD,Wang2022aggregate3DFD}, and (iii) its particular suitability to sets of points or volume elements~\cite{Zhao2016MRI3DFD}. The box-counting dimension $D_{box}$ could be computed by breaking a 3D object into increasingly smaller voxels with side length $\epsilon$, and counting the number of boxes containing patterns of interest $N(\epsilon)$ at each scale. \par

Given that surfaces of atomistic objects are not expected to be strictly self-similar, we can model a given surface as a subset of the 3D vector space ($\mathbb{R}^3$), and compute its $D_{box}$ to quantify its complexity. However, the 3D objects in previous studies (particularly in medical, ecology, geography, and materials research) are often represented as 3D images, and the ``boxes'' are either pixels or voxels~\cite{Wang2015sapphire3DFD,Wu2010MRI3DFD,Pashminehazar2019agglomerate3DFD,Zhao2016MRI3DFD,Feranie2011rock3DFD,Kobayashi1994electricTree3DFD,Tang2006fatCrystal3DFD,Tang2008fatCrystal3DFD,Esteban2009MRI3DFD,Bayarri2010MRI3DFD,Vahedi2011floc3DFD,Helmberger2014lung3DFD,Krohn2019MRI3DFD,Schurch2019tree3DFD,Wu2019coal3DFD,Xia2019rock3DFD,Li2022rock3DFD}. This limits the precision of $D_{box}$ as it depends on the resolution of the images. In the case of atomistic objects, high-resolution 3D images consisting of voxels impose a stringent demand on memory storage, which renders this representation inappropriate for scenarios that involve many conformations of the objects, particularly high-throughput studies. On the other hand, if the binary images are not of adequately high resolution, the granularity of the surface details will be compromised and the representation will be inaccurate. A balance between accuracy and computational efficiency must be struck. Researchers need to be able to represent the surface of the atomistic objects exactly, ideally taking atomic coordinates directly as input, to obtain the most accurate estimation of the overall fractal dimension while minimising memory storage and computational time. \par

In this work, we propose two approaches to estimate the fractal dimensions of atomistic objects: box-counting algorithms on surfaces approximated as voxelised point clouds, or objects represented in precise mathematical form. We introduce Sphractal as an open source, platform agnostic Python package for these calculations by describing its implementation, and demonstrate its utility by calculating the $D_{box}$ of a set of palladium nanoparticles. \par

\section{Methodology}
\label{sec:methods}
Given a set of atomic coordinates, with each atom represented as a sphere with a particular radius, the atomistic surface is formed by the outward-facing non-overlapping parts of the atoms lying on the surface. To compute the $D_{box}$ of such surfaces, we perform the steps outlined in Algorithm~\ref{alg:calcBoxCountDim}, which are explained below: \par

\begin{enumerate}
    \item Identify the neighbouring atoms of each atom $a_i$ using a cutoff set to be either the atomic or metallic radius $R_i$ of the atom multiplied by a constant to account for the fluctuations in the length of the chemical bond between them. The atomic and metallic radii are obtained from the work of Clementi, Raimondi, and Reinhardt~\cite{Clementi2004atomRad} and from the work of Greenwood and Earnshaw~\cite{Greenwood2012metalRad}, respectively. Users may specify the preferred radius type and value of the multiplier.
    \item Identify the surface atoms using either the convex hull method, number of neighbours criterion, or the alpha shape algorithm~\cite{Edelsbrunner1994alphaShape}. The convex hull method is incapable of detecting concave surfaces and therefore prone to false negatives (missing actual surface atoms). The simplicity of the number of neighbour criterion (any atom with less than a specified number of neighbour is considered as surface atom) renders it much less computationally expensive than the alternatives, but also very prone to false positives (misidentify inner atoms as surface). The default recommendation for \revision{users} is to use the alpha shape algorithm with the value of $\alpha$ set to the smallest atomic diameter, as an overly large or small $\alpha$ will also result in false positives and negatives. 
    \item Conduct box-counting on the surface of the atomistic object represented as either voxelised point cloud, or mathematically precise surface. This step will be illustrated further in Sections~\ref{sec:pointCloud} and \ref{sec:exactSurf}.
    \item Compute $D_{box}$, which is the slope of the log-log plot of change in box counts with respect to the scaling of the box lengths, along with the coefficient of determination $R^2$ and the confidence intervals at the confidence level specified by \revision{users}. The range of box lengths from which the slope is to be calculated is automatically selected since the range in which naturally occurring atomistic objects (such as nanoparticles or proteins) exhibit self-similarity is generally unknown \emph{a priori}. Here we search for a range where the $R^2$ of the linear regression fitted to the data is maximised~\cite{Wu2010MRI3DFD,Zhao2016MRI3DFD,Marzi2018FDbrain}. The range is determined by successively removing the last point while $R^2$ continues to increase, and then successively removing the first point until $R^2$ decreases~\cite{Helmberger2014lung3DFD} (with the retention of a minimum number of points specified by \revision{users}). The slope calculation procedure is described in Algorithm~\ref{alg:calcSlope}.
\end{enumerate}

\begin{algorithm}
    \caption{Pseudocode for estimation of fractal dimension of the surface of an atomistic object via box-counting technique.}
    \label{alg:calcBoxCountDim}
    \begin{flushleft}
        \textbf{Input}: \\ 
        $\protect\overrightarrow{A}: [a_1,a_2,...,a_{N_a}]$, $a_i$ being atom $i$ with Cartesian coordinates $\protect\overrightarrow{r}_i: (x_i,y_i,z_i) \in \mathbb{R}^3$, radius $R_i \in \mathbb{R}_{>0}$, and surface atom indicator $s_i \in \{0,1\}$ of an object consisting of $N_a \in \mathbb{Z}$ atoms \\
        $n_{sample\_min}:$ minimum number of samples to retain for the calculation of the slope \\
        \textbf{Output}: \\
        $D_{box}:$ box-counting dimension of the surface of an object composed of spheres
    \end{flushleft}
    \begin{algorithmic}[1] 
        \Procedure{calcBoxCountDim}{$\protect\overrightarrow{A},n_{sample\_min}$}
            \State Generate nearest neighbours list for $\protect\overrightarrow{A}$
            \State Identify and mark the surface atoms
            \State Compute $\protect\overrightarrow{l},\protect\overrightarrow{N}$ by running box-counting algorithm on the surface represented as either voxelised point cloud or mathematically precise object
            \State $D_{box} \gets$ \Call{calcSlope}{$\protect\overrightarrow{l},\protect\overrightarrow{N}$,$n_{sample\_min}$}
            \State \textbf{return} $D_{box}$
        \EndProcedure
    \end{algorithmic}
\end{algorithm}

\begin{algorithm}
    \caption{Pseudocode for calculation of the slope of log-log plot of box counts over box lengths.}
    \label{alg:calcSlope}
    \begin{flushleft}
        \textbf{Input}: \\
        $\protect\overrightarrow{l}: [l_1,l_2,...]$, $l_i$ being box length $i$ \\ 
        $\protect\overrightarrow{N}: [N_1,N_2,...]$, $N_i$ being number of boxes with box length $i$ that cover the surface of interest \\
        $n_{sample\_min}:$ minimum number of samples to retain for the calculation of the slope \\
        \textbf{Output}: \\
        $D_{box}:$ box-counting dimension of the surface of the given object \\
        $CI:$ confidence interval of estimated slope \\
        $R^2:$ coefficient of determination of the linear regression \\
        $l_{min}:$ smallest box length \\
        $l_{max}:$ largest box length
    \end{flushleft}
    \begin{algorithmic}[1]
        \Procedure{calcSlope}{$\protect\overrightarrow{l},\protect\overrightarrow{N}$,$n_{sample\_min}$}
            \State $p0 \gets 0$  \Comment{Index of first data point}
            \State $p1 \gets length(\protect\overrightarrow{N})$   \Comment{Index of last data point}
            \State Initialise $D_{box}$, $CI$, $R^2$, $l_{min}$, $l_{max}$
            \State $removeFrom \gets back$  \Comment{Indicator for direction to remove points from}
            \While{$length(\protect\overrightarrow{N}) \ge n_{sample\_min}$}
                \State $\protect\overrightarrow{l} \gets \protect\overrightarrow{l}[p0:p1]$
                \State $\protect\overrightarrow{N} \gets \protect\overrightarrow{N}[p0:p1]$
                \State Fit an ordinary least squares regressor to $\left(\text{log}(\protect\overrightarrow{l}),\text{log}(\protect\overrightarrow{N})\right)$
                \State Compute $D_{box}$, $CI$, and $R^2$ from the linear regression slope
                \State $l_{min} \gets \protect\overrightarrow{l}[0]$
                \State $l_{max} \gets \protect\overrightarrow{l}[-1]$
                \State $R^2_{prev} \gets R^2$ from previous iteration
                \If{$removeFrom == back$}
                    \If{$R^2<R^2_{prev}$}
                        \State $removeFrom \gets front$  \Comment{Start removing from the first point next}
                    \Else
                        \State $p1 \gets p1-1$
                    \EndIf
                \Else
                    \If{$R^2<R^2_{prev}$}
                        \State \textbf{return} $D_{box}$, $CI$, $R^2$, $l_{min}$, $l_{max}$ from previous iteration
                    \Else
                        \State $p0 \gets p0+1$
                    \EndIf
                \EndIf
                \State Store $D_{box}$, $CI$, $R^2$, $l_{min}$, $l_{max}$ for next iteration
            \EndWhile
            \State \textbf{return} $D_{box}$, $CI$, $R^2$, $l_{min}$, $l_{max}$
        \EndProcedure
    \end{algorithmic}
\end{algorithm}

\subsection{Voxelised Point Cloud Representation}
\label{sec:pointCloud}

\begin{figure} \centering
    \sidesubfloat[]{\includegraphics[width=0.35\textwidth]{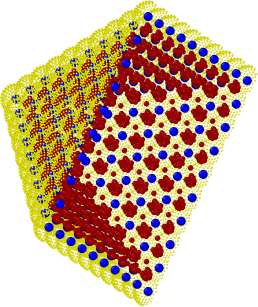}\label{fig:exampleVXpointCloudSliced}} 
    \hspace{0.05\textwidth}
    \sidesubfloat[]{\includegraphics[width=0.35\textwidth]{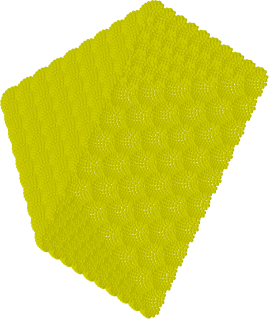}\label{fig:exampleVXvoxelsSliced}}
    \caption{Slice through illustrations of the (a) point cloud approximation and (b) voxelised representation of the surface of an octahedral nanoparticle. In the left figure, the yellow and red points lie on the outer surface (which are of interest) and inner surface (which are removed in this case, but could be retained if desired), respectively; the blue points are the centers of the surface atoms around which the yellow and red points were generated. In the right figure, the yellow boxes correspond to the voxels occupied by the yellow points in the left figure.}
    \label{fig:egVX}
\end{figure}

To apply a typical box-counting algorithm to an atomistic surface, the surface needs to be represented as an image. A binary 3D voxelised image is an appropriate representation, but direct transformation from the coordinates of centre of atoms into such an image is difficult. A more feasible approach is to first approximate the surface as a point cloud. This can be achieved by generating evenly spaced points around each atom and removing the points that overlap with other atoms or fall within the inner surface (see Algorithm~\ref{alg:isInnerSide}). Users can choose to retain the inner surface points if desired. The points around each atom are obtained using the Fibonacci lattice~\cite{Gonzalez2009FibonacciSphere}. The resulting point cloud can then be transformed into a voxelised representation of the surface in a straightforward manner. Figure~\ref{fig:egVX} depicts an example of the point cloud approximation of an atomistic surface and its corresponding voxelised representation. \par

\begin{algorithm}
    \caption{Pseudocode to examine whether a given point or box lies on the inner side of an object.}
    \label{alg:isInnerSide}
    \begin{flushleft}
        \textbf{Input}: \\
        $\protect\overrightarrow{p}:$ coordinate of a point or centre of a box in $\mathbb{R}^3$ \\ 
        $\protect\overrightarrow{q}:$ coordinate of the centre of an atom in $\mathbb{R}^3$ \\ 
        \textbf{Output}: \\
        $isInner:$ boolean value of whether the point or box lies on the inner side
    \end{flushleft}
    \begin{algorithmic}[1]
        \Procedure{isInnerSide}{$\protect\overrightarrow{p}, \protect\overrightarrow{q}$}
            \State Classify the neighbours of atom with coordinate $\protect\overrightarrow{q}$ into surface and inner atoms
            \State Compute average coordinate of inner atoms $\protect\overrightarrow{q_I}$
            \State Find coordinates of closest surface atom pairs $\protect\overrightarrow{q_2},\protect\overrightarrow{q_3}$ from $\protect\overrightarrow{p}$, that are also a neighbour of each other
            \State $\protect\overrightarrow{n} \gets (\protect\overrightarrow{q_2} - \protect\overrightarrow{q}) \times (\protect\overrightarrow{q_3} - \protect\overrightarrow{q})$  \Comment{Normal of the plane formed by the surface atoms}
            \State $\protect\overrightarrow{v}_1 \gets \protect\overrightarrow{q_I}-\protect\overrightarrow{q}$  \Comment{From atom centre to average of inner atoms}
            \State $\protect\overrightarrow{v}_2 \gets \protect\overrightarrow{p}-\protect\overrightarrow{q}$  \Comment{From atom centre to the surface point or box centre}
            \If{$(\protect\overrightarrow{n} \cdot \protect\overrightarrow{v}_1) \circ (\protect\overrightarrow{n} \cdot \protect\overrightarrow{v}_2) < 0$}  \Comment{If $\protect\overrightarrow{v}_1$ and $\protect\overrightarrow{v}_2$ point in different directions}
                \State $isInner \gets False$
            \Else  \Comment{If $\protect\overrightarrow{v}_1$ and $\protect\overrightarrow{v}_2$ point in the same direction}
                \State $isInner \gets True$
            \EndIf
            \State \textbf{return} $isInner$
        \EndProcedure
    \end{algorithmic}
\end{algorithm}

We have incorporated the most efficient box-counting algorithm in the literature to date~\cite{Miras2023GPUFD} into Sphractal. Readers are directed to the original publication for details on this algorithm and its implementation. A drawback of this approach is that the original work traded off the ability to specify the starting and ending box lengths, and the gap of reductions in box lengths for higher speed of computation. However, the flexibility of defining these parameters may be useful to study atomistic objects, which can only display fractal behaviour over a relatively small range. This issue is addressed in the algorithm we developed in this work. \par

\subsection{Mathematically Exact Surface Representation}
\label{sec:exactSurf}

Alternatively, the box-counting algorithm can be adapted to check whether a given box covers the surface of interest, which could also optionally include the inner surface apart from the outer surface. For each box length $l$, we aim to count the number of boxes that cover the non-overlapping parts of the spherical surface of the atoms with element-dependent radius $R$. For each atom $a$, $2\lceil\frac{R+l}{l}\rceil+1$ boxes around it are examined, taking into account the uncertainty in the position of $a$ within the box. A box is counted as covering the surface of interest if it satisfies the following criteria (the inner surface could be included by considering all atoms as surface atoms and ignoring the last criterion: \par 

\begin{enumerate}[(i)]
    \item its nearest and farthest points from any surface atom (computed via Algorithm~\ref{alg:getNearFarBoxCoord}) falls within and without the radius of the atom respectively, 
    \item it is not completely encompassed within the radius of any atom, and
    \item its centre does not fall on the same side as the inner atoms relative to the plane formed by the closest surface atoms (as assessed using Algorithm~\ref{alg:isInnerSide}).
\end{enumerate}

\begin{algorithm}
    \caption{Pseudocode for computing the nearest and furthest coordinates of a box from an atom along an axis.}
    \label{alg:getNearFarBoxCoord}
    \begin{flushleft}
        \textbf{Input}: \\
        $O:$ coordinate of the atom along an axis \\ 
        $b_{min}:$ lower boundary of the box along an axis \\ 
        $b_{max}:$ upper boundary of the box along an axis \\ 
        $l:$ length of the box \\
        \textbf{Output}: \\
        $r_{near}:$ nearest coordinate of the box from the atom along an axis \\ 
        $r_{far}:$ farthest coordinate of the box from the atom along an axis
    \end{flushleft}
    \begin{algorithmic}[1]
        \Procedure{getNearFarBoxCoord}{$O,b_{min},b_{max},l$}
            \If{$O<b_{min}$}  \Comment{If the whole box is to the right of the atom}
                \State $r_{near} \gets b_{min}$
                \State $r_{far} \gets b_{max}$
            \ElsIf{$O>b_{max}$}  \Comment{If the whole box is to the left of the atom}
                \State $r_{near} \gets b_{max}$
                \State $r_{far} \gets b_{min}$
            \Else  \Comment{If the atom falls within the box}
                \State $r_{near} \gets O$  \Comment{Nearest coordinate is the coordinate of the atom itself}
                \If{$b_{max}-O < l/2$}  \Comment{If the atom is closer to the right boundary}
                    \State $r_{far} \gets b_{min}$
                \Else
                    \State $r_{far} \gets b_{max}$
                \EndIf
            \EndIf
            \State \textbf{return} $r_{near},r_{far}$
        \EndProcedure
    \end{algorithmic}
\end{algorithm}

\begin{figure}[htbp]  \centering
\includegraphics[width=0.25\textwidth]{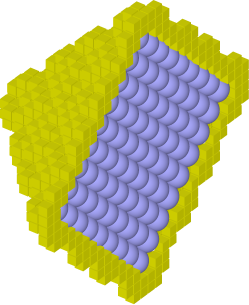} \hspace*{8mm}
\includegraphics[width=0.25\textwidth]{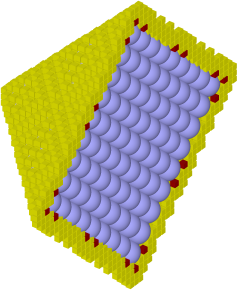} \hspace*{8mm}
\includegraphics[width=0.25\textwidth]{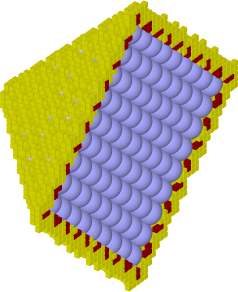} \\
\includegraphics[width=0.25\textwidth]{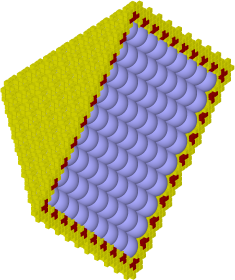} \hspace*{8mm}
\includegraphics[width=0.25\textwidth]{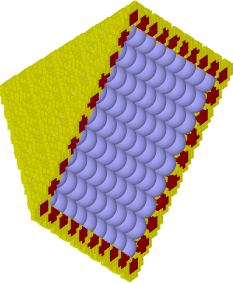} \hspace*{8mm}
\includegraphics[width=0.25\textwidth]{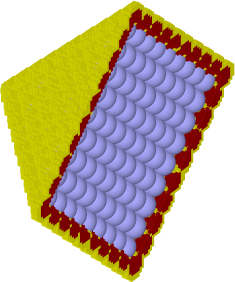} \\
\caption{Example output coordinates of the boxes examined over the mathematically exact surface representation of an octahedral palladium nanoparticle with increasingly smaller boxes. Each light blue sphere represents a palladium atom, while the yellow and red boxes correspond to the boxes that cover and do not cover the outer surface, respectively.}
\label{fig:egEX}
\end{figure}

The box-counting procedure is described in Algorithm~\ref{alg:exactBoxCnt}. Figure~\ref{fig:egEX} shows some example output coordinates of the boxes examined over a metal nanoparticle surface. The algorithm allows us to capture the whole atomistic surface without any approximation, hence removing the dependence of the resulting box-counting dimension on the resolution of the surface representation. 
That said, the computed dimension is dependent upon. \par 

\begin{algorithm}
    \caption{Pseudocode for counting the number of boxes covering the outer surface of an atomistic object for a particular box length.}
    \label{alg:exactBoxCnt}
    \begin{flushleft}
        \textbf{Input}: \\
         $\protect\overrightarrow{F}: [f_1,f_2,...]$, $f_i$ being surface atom $i$ with Cartesian coordinates $\protect\overrightarrow{r_i}: (x_i,y_i,z_i) \in \mathbb{R}^3$, and radius $R_i \in \mathbb{R}_{>0}$ \\
        $l:$ length of boxes \\
        \textbf{Output}: \\
        $S:$ identifiers of boxes covering the surface of interest
    \end{flushleft}
    \begin{algorithmic}[1]
        \Procedure{exactBoxCount}{$\protect\overrightarrow{F},l$}
            \State $S \gets \emptyset$
            \State $B \gets \emptyset$  \Comment{Boxes covering inner part of the object}
            \For{$f_i \in \protect\overrightarrow{F}$}
                \State Identify $\protect\overrightarrow{i}: (i_x,i_y,i_z)$  \Comment{Index of the box that ${f}_i$ falls within}
                \State $n_{scan} \gets \lceil(R_i+l) / l\rceil$  \Comment{Number of boxes to scan around $\protect\overrightarrow{i}$}
                \For{$i_x-n_{scan} < j_x < i_x+n_{scan}$}
                    \For{$i_y-n_{scan} < j_y < i_y+n_{scan}$}
                        \For{$i_z-n_{scan} < j_z < i_z+n_{scan}$}
                            \State Compute coordinates of box centre $\protect\overrightarrow{p}$
                            \If{\Call{isInnerSide}{$\protect\overrightarrow{p},\protect\overrightarrow{r_i}$}}
                                \State Continue
                            \EndIf
                            
                            \State Compute boundaries of $\protect\overrightarrow{j}: (b_{xmin},b_{ymin},b_{zmin},b_{xmax},b_{ymax},b_{zmax})$
                            \State $x_{near},x_{far} \gets$ \Call{getNearFarCoord}{$x_i,b_{xmin},b_{xmax},l$}
                            \State $y_{near},y_{far} \gets$ \Call{getNearFarCoord}{$y_i,b_{ymin},b_{ymax},l$}
                            \State $z_{near},z_{far} \gets$ \Call{getNearFarCoord}{$z_i,b_{zmin},b_{zmax},l$}
                            \State Compute $d_{near},d_{far}$ from $x_{near},y_{near},z_{near}$ and $x_{far},y_{far},z_{far}$
                            
                            \If{$d_{near} < R_i < d_{far}$}
                                \State $S \gets S \cup \protect\overrightarrow{j}$
                            \ElsIf{$d_{far} < R_i$}
                                \State $B \gets B \cup \protect\overrightarrow{j}$
                            \EndIf
                        \EndFor
                    \EndFor
                \EndFor
            \EndFor
            \State $S \gets S \setminus B$
            \State \textbf{return} $S$
        \EndProcedure
    \end{algorithmic}
\end{algorithm}

Users are given the ability of tuning a few essential parameters, including the range and sampling interval of the box lengths. An appropriate range of box lengths is important to mitigate biases in the results~\cite{Zhao2016MRI3DFD}. The box counts obtained from larger boxes deviate from the power law observed when the atomistic object demonstrates self-similarity (being far away from the linear slope formed by the rest of the box counts). On the other hand, the estimated fractal dimension approaches $2.0$ as the box length reduces. This is because an atomistic surface essentially conforms to a 2D surface as the capturing of the growth in spherical surfaces dominates the capturing of the characteristic details of the `grooves' formed between atoms. An overly small or large sampling interval of box lengths also leads to high sensitivity of the algorithm to noises and results in high uncertainty in the slope estimated. \par 

As most \revision{users} will likely be interested in capturing the complexity at the atomic scale, unnecessary computations could be avoided by confining the range of box lengths to values around the minimum atomic radius $R_{min}$. To achieve a balance between computational cost and uncertainty in slope estimation, the largest and smallest box lengths are set to $R_{min}$ and $R_{min} / 4$, respectively. The default value for sampling size is 10 evenly-spaced box lengths on the logarithmic scale. Users are advised to conduct some tuning of these parameters prior to production runs. \par

\section{Software Details}
\label{sec:softwareDetails}

\subsection{Implementation}
\label{sec:implementation}

Sphractal is distributed under an open source software license (MIT). We implemented Sphractal in Python using state-of-the-art packages for scientific computing, visualisation, and statistical modelling, namely \texttt{NumPy}~\cite{Harris2020NumPy}, \texttt{SciPy}~\cite{Virtanen2020SciPy}, \texttt{Matplotlib}~\cite{Hunter2007Matplotlib}, and \texttt{statsmodels}~\cite{Seabold2010statsmodels}. Additionally, the \texttt{Numba} compiler~\cite{Lam2015Numba} was used to speed up the algorithm via just-in-time compilation. The point cloud generation step for voxelised point cloud surface approximation and box-counting step for mathematically exact surface representation were further optimised via parallelisations. The complete implementation of Sphractal's functionalities, its documentation, and examples of its applications are publicly available at \url{https://github.com/jon-ting/sphractal}. \par 

\begin{figure}[hp!]
\centering
\includegraphics[width=0.9\textwidth]{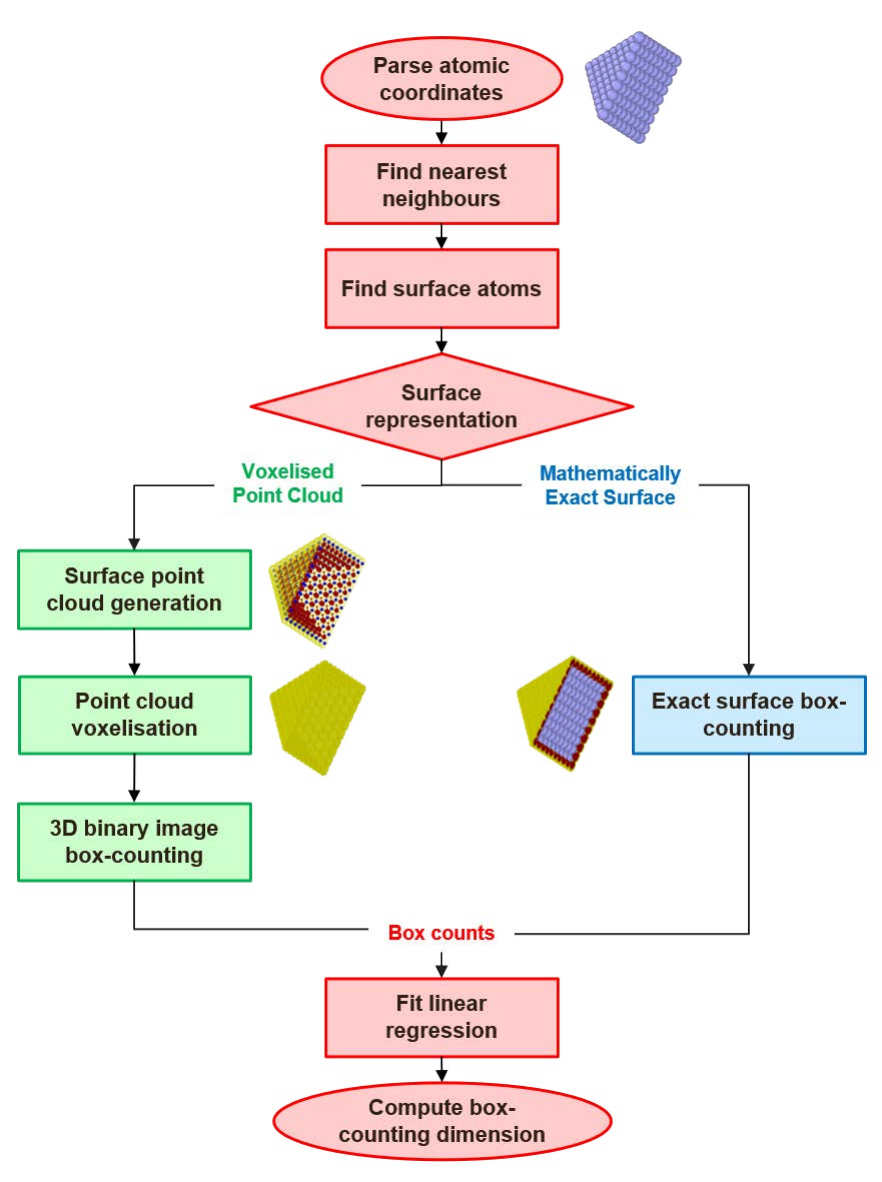}
\caption{Overview of Sphractal workflow. The steps coloured in red signify a common pipeline, while those coloured in green and blue correspond to the steps required only for the voxelised point cloud representation, and mathematically exact representation, respectively.}
\label{fig:sphractalWorkflow} 
\end{figure}

\subsection{Utility}
\label{sec:utility}

The workflow of Sphractal is illustrated in Figure~\ref{fig:sphractalWorkflow}. The main modules of Sphractal are described below \revision{and demonstrated in Listing~\ref{lst:sphractalDemo}}: \par

\begin{enumerate}
    \item \texttt{constants} contains the constant values used by Sphractal, such as the atomic and metallic radii of all elements and configurations of output figures. These values could then be inspected.
    \item \texttt{datasets} provides access to the atomic coordinates used for benchmarking, scaling tests, complexity analysis, and validation of the algorithms (presented in Section~\ref{sec:performanceAnalysis}). 
    \item \texttt{utils} provides functionalities for parsing of the coordinate file, building of neighbour lists for each atom, and identification of surface atoms. 
    \item \texttt{surfVoxel} generates the point cloud, performs the subsequent voxelisation, and applies the state-of-the-art box-counting algorithm for 3D binary images to the voxel representation of the surface.
    \item \texttt{surfExact} hosts the codes for the box-counting algorithm adapted for the mathematically exact surface representation.
    \item \texttt{boxCnt} provides an interface for box-counting to be performed on the specified representation(s) of a given surface, and outputs its box-counting dimension along with the coefficient of determination, confidence intervals, and range of box lengths used to compute the dimension. It also allows the results to be visualised. 
\end{enumerate}

\begin{mypython}[caption={Demonstration of the modules in Sphractal.}, label={lst:sphractalDemo}] 
# Demonstration of the constants module
from sphractal.constants import ATOMIC_RAD_DICT, METALLIC_RAD_DICT, PLT_PARAMS

print(ATOMIC_RAD_DICT['Au'])
print(METALLIC_RAD_DICT['Au'])
print(PLT_PARAMS['notebook'])

# Demonstration of the datasets module
from sphractal.datasets import getExampleDataPath, getStrongScalingDataPath,
                               getWeakScalingDataPaths, getValidationDataPath, 
                               getCaseStudyDataPaths

inpFile = getExampleDataPath()
strongScalingXYZ = getStrongScalingDataPath()
weakScalingXYZs = getWeakScalingDataPaths()
validationXYZ = getValidationDataPath()
caseStudyXYZ = getCaseStudyDataPaths()

# Demonstration of the utils module
from sphractal.utils import readInp, findNN, findSurf

eles, rads, xyzs, maxRange, minXYZ, maxXYZ = readInp(inpFile)
maxRad, radMult = rads.max(), 1.2
neighs, bondLens = findNN(rads, xyzs, minXYZ, maxXYZ, maxRad, radMult)
surfs = findSurf(xyzs, neighs)

# Demonstration of the surfVoxel module
from sphractal.surfVoxel import voxelBoxCnts

scalesVX, countsVX = voxelBoxCnts(eles, rads, surfs, xyzs, neighs)

# Demonstration of the surfExact module
from sphractal.surfExact import exactBoxCnts

minMaxBoxLens = (rads.min() * 0.25, rads.min())
scalesEX, countsEX = exactBoxCnts(eles, rads, surfs, xyzs, neighs,
                                  maxRange, minMaxBoxLens, minXYZ)

# Demonstration of the boxCnt module
from sphractal.boxCnt import findSlope, runBoxCnt

r2VX, dBoxVX, ciVX, boxLensVX = findSlope(scalesVX, countsVX)
r2EX, dBoxEX, ciEX, boxLensEX = findSlope(scalesEX, countsEX)

r2VX, dBoxVX, ciVX, boxLensVX, r2EX, dBoxEX, ciEX, boxLensEX = runBoxCnt(inpFile)
\end{mypython}

The optional arguments are omitted from \revision{Listing~\ref{lst:sphractalDemo}} to simplify the demonstrations. The simplest usage of Sphractal requires only the calling of a single function \texttt{runBoxCnt()}, with the path to the file containing the atomic coordinates of an atomistic object as input argument. As the function \texttt{runBoxCnt()} is used to generate all data for Sections~\ref{sec:validation}, \ref{sec:performanceAnalysis}, and \ref{sec:caseStudy}, an important subset of its parameters are explained in detail under Table~\ref{tab:runBoxCntParams}. \par

\begin{table}
\centering
 \begin{tabular}[htbp]{@{}cccl@{}}
 \hline
 Parameter & Data Type & Default Value & Description \\
 \hline
 inpFilePath & String & - & Path to xyz file containing Cartesian coordinates \\ & & & of a set of atoms. \\
 radType & String & \texttt{`atomic'} & Type of radii to use for the atoms, one of \\ & & & \texttt{\{`atomic', `metallic'\}}. \\
 radMult & Numeric & \texttt{1.2} & Multiplier to the radii of atoms to identify their \\ & & & neighbouring atoms. \\
 findSurfAlg & String & \texttt{`alphaShape'} & Algorithm to identify the surface atoms, one of \\ & & & \texttt{\{`alphaShape', `convexHull', `numNeigh'\}}. \\
 alphaMult & Numeric & \texttt{2.0} & Multiplier to the minimum radius to decide $\alpha$ \\ & & & value for the alpha shape algorithm. \\
 trimLen & Boolean & \texttt{True} & Whether to remove the box counts obtained \\ & & & using boxes of extreme sizes. \\
 minSample & Integer & \texttt{6} & Minimum number of data points to retain for \\ & & & slope estimation from linear regression fitting. \\
 confLvl & Numeric & \texttt{95} & Confidence level of confidence intervals (\%). \\
 rmInSurf & Boolean & \texttt{True} & Whether to remove the inner surface points. \\
 voxelSurf & Boolean & \texttt{True} & Whether to represent the surface as voxelised \\ & & & point clouds. \\
 numPoints & Integer & \texttt{10000} & Number of surface points to generate around \\ & & & each atom. \\
 gridNum & Integer & \texttt{1024} & Resolution of the 3D binary image. \\
 exactSurf & Boolean & \texttt{True} & Whether to represent the surface in a \\ & & & mathematically exact manner. \\
 minLenMult & Numeric & \texttt{0.25} & Multiplier to the minimum radius to determine \\ & & & the minimum box length. \\
 maxLenMult & Numeric & \texttt{1.0} & Multiplier to the maximum radius to determine \\ & & & the maximum box length. \\
 numCPUs & Integer & \texttt{8} & Number of cores to be used for parallelisation. \\
 numBoxLen & Integer & \texttt{10} & Number of box length samples for the collection \\ & & & of the box count data, spaced evenly on \\ & & & logarithmic scale. \\ 
 \hline
 \end{tabular}
\caption{Explanation of important parameters of the \texttt{runBoxCnt()} function.}
\label{tab:runBoxCntParams}
\end{table}

Sphractal can optionally generate a few output files that allow the results to be visualised. For the voxel representation of the surface, the coordinates of the surface points and of the voxels corresponding to the points lying on the surface of interest could be stored as separate files. Similarly, the coordinates of the boxes covering the surface could be stored for the mathematically exact surface representation. Examples of these coordinate files were shown in Figures~\ref{fig:egVX} and \ref{fig:egEX}. The figures in this work were generated using the Open Visualisation Tool~\cite{Stukowski2009OVITO}. \par

\subsection{Installation}
\label{sec:installation}
The package can be installed from Python Package Index and conda-forge as illustrated in Listing~\ref{lst:installSphractal}. To enable the application of the state-of-the-art box-counting algorithm to the voxelised point cloud surface representation, \revision{users} also need to install a C++ program adapted from the work of Juan Ruiz de Miras et al.~\cite{Miras2023GPUFD}, as instructed at \url{https://github.com/jon-ting/fastbc} and illustrated by Listing~\ref{lst:installFastBC}. \par

\lstset{
    numbers=left,
    stepnumber=1,
    firstnumber=1,
    numberfirstline=true
}

\begin{terminal}[caption={Installation of Sphractal through either line 1 or line 2.}, label={lst:installSphractal}] 
pip install sphractal
conda install -c conda-forge sphractal
\end{terminal}

\begin{terminal}[caption={Compilation of the 3D binary image box-counting program through either line 3 or line 4 (line 4 requires the corresponding C++ compiler). \revision{PATH\_TO\_FASTBC} in line 5 needs to be replaced with the absolute path to the compiled \revision{file}.},label=
{lst:installFastBC}]
git clone https://github.com/jon-ting/fastbc.git
cd fastbc
g++ 3DbinImBCcpu.cpp bcCPU.cpp -o 3DbinImBCcpu
nvcc -O3 3DbinImBCgpu.cpp bcCUDA3D.cu -o 3DbinImBCgpu
export FASTBC=PATH_TO_FASTBC
\end{terminal}

\section{Validation}
\label{sec:validation}

\begin{figure}[htbp]  \centering
\sidesubfloat[]{\includegraphics[width=0.45\textwidth]{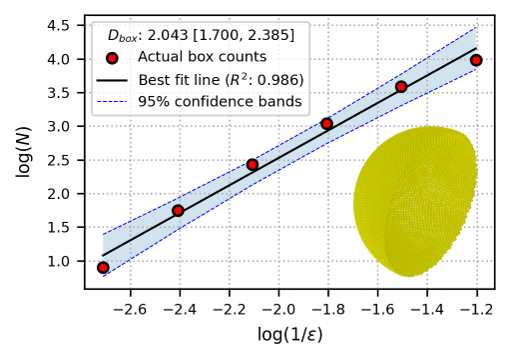}\label{fig:DboxSingleAtomVX}}
\sidesubfloat[]{\includegraphics[width=0.45\textwidth]{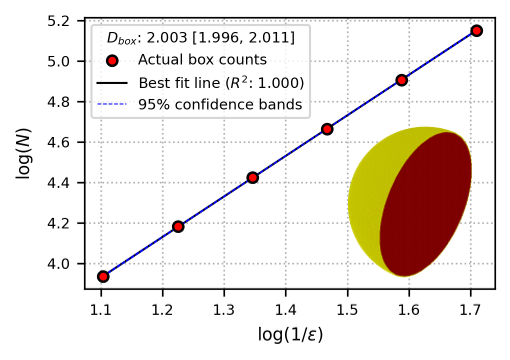}\label{fig:DboxSingleAtomEX}} \\
\caption{Plots showing ordinary least squares best fit lines to box-counting data collected from (a) voxelised point cloud and (b) mathematically exact surface representations. The generated point cloud and boxes examined were attached to the former and latter representations, respectively.}
\label{fig:singleAtomValidation}
\end{figure}

To validate the correctness of the implementations, the algorithms are applied to \revision{a mathematical surface with known fractal dimension. Here, for simplicity, a single palladium atom (corresponding to the surface of a sphere with fractal dimension of $2$) has been chosen as the test case.} As shown in Figure~\ref{fig:singleAtomValidation}, the $D_{box}$ values calculated from the box-counting on both surface representations closely match the expected value. The relevant (and non-default) parameter values used to obtain the results were chosen to (i) accommodate for the much smaller system size compared to the typical settings and (ii) yield results with lower uncertainty: \texttt{rmInSurf=False}, \texttt{numPoints=10000}, \texttt{minLenMult=0.005}, \texttt{numBoxLen=20}. \par

\section{Performance Analysis}
\label{sec:performanceAnalysis}

The time complexity of the implemented algorithms was analysed via some tests conducted on simulated nanoparticles with diameters ranging from 10 \AA\ (43 atoms) to 123 \AA\ (66283 atoms). The atomic coordinates of the nanoparticles were generated using publicly available Atomic Simulation Environment tools~\cite{Larsen2017ASE} and are accessible from within Sphractal under the \texttt{datasets} module as demonstrated in \revision{Listing~\ref{lst:sphractalDemo}}. The tests were run using \revision{12 cores on a high-performance computing cluster.} The technical specifications of the machine used are enumerated under Table~\ref{tab:techSpecs}. Readers are directed to the Supporting Information for results obtained from scalability tests. \par

\begin{table}[htbp]
\centering
 \begin{tabular}{cc} 
 \hline
 Aspects & High-performance computing cluster \\
 \hline
 Operating system & Rocky Linux 8 x86\_64 \\ 
 CPU model & Intel Xeon Platinum 8268 (Cascade Lake) 2.9 GHz \\
 CPUs per node & 2 (each with 2 NUMA nodes) \\
 Cores per CPU & 24 (12 cores per NUMA node) \\
 RAM per node (GB) & 384 (96 GB local RAM per NUMA node) \\
 Interconnect & Mellanox HDR 200G InfiniBand in \\ & Dragonfly+ network topology 200 Gb/s \\
 \revision{GPU model} & \revision{Nvidia Tesla Volta V100-SXM2-32GB} \\
 \hline
 \end{tabular}
\caption{Technical specifications of the machine on which the complexity and scaling tests were carried out.}
\label{tab:techSpecs}
\end{table}

The tests were run using parameter values that are expected to resemble a production run setting. These values are mostly the default values in Sphractal, except \texttt{saveFig=True}, \texttt{verbose=True}, and \texttt{numPoints=300}. The run times reported were averages of 30 runs after outlier removal. The box plots of the run time distributions for all tests are included in the Supporting Information. \par

Since the box-counting algorithm for the voxel representation of atomistic surface relies heavily on the algorithm published by Ruiz de Miras et al.~\cite{Miras2023GPUFD}, readers are directed to the original publication for details on the performances of the various implementations of the algorithm (sequential CPU, parallelised CPU, and GPU-accelerated). \revision{The GPU-accelerated implementation of the algorithm was employed in the following tests.} \par 

\begin{figure}[htbp]  \centering
\sidesubfloat[]{\includegraphics[width=0.45\textwidth]{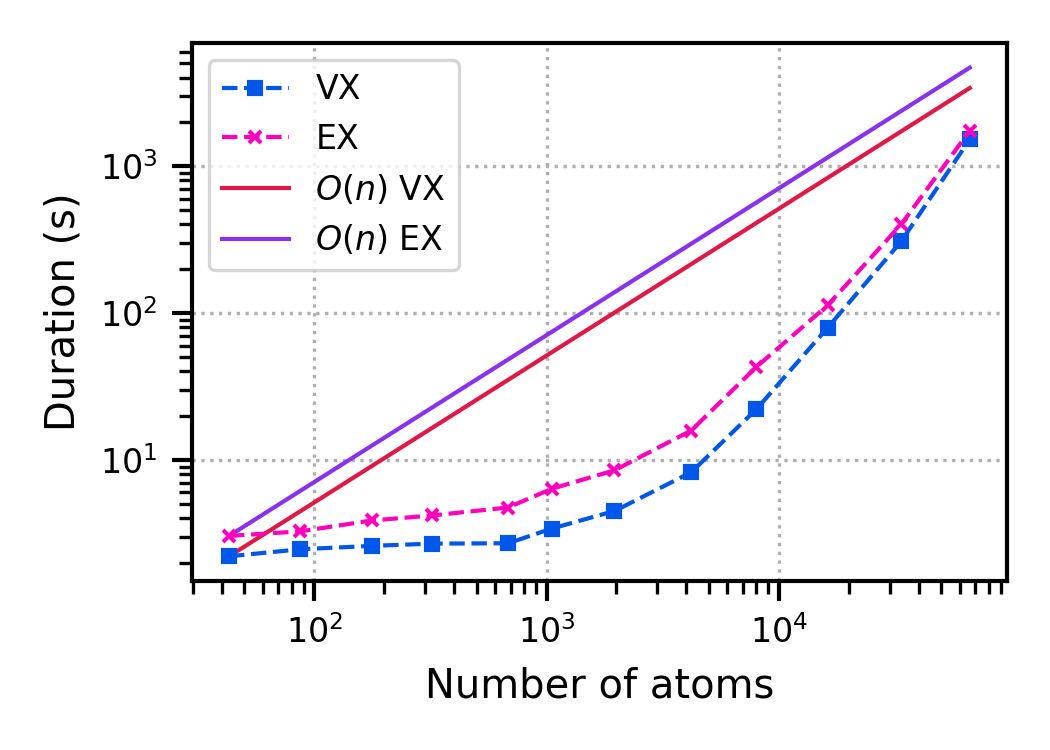}\label{fig:TC_VXEX_loglog}}
\sidesubfloat[]{\includegraphics[width=0.45\textwidth]{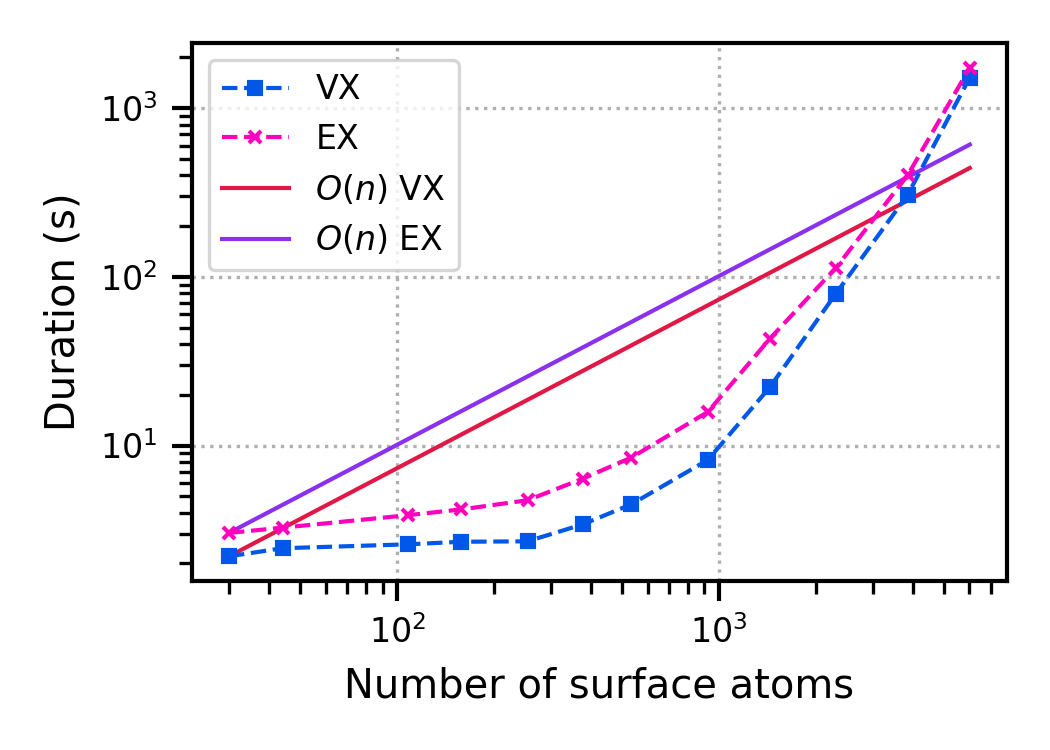}\label{fig:TC_VXEX_loglog_satom}} \\
\caption{\revision{Run times for box-counting of the voxelised point cloud and mathematically exact surface representations in logarithmic scales, as a function of (a) total number of atoms and (b) total number of surface atoms.}}
\label{fig:loglogTimeComplexityPlots}
\end{figure}

Figure~\ref{fig:loglogTimeComplexityPlots} shows the growth in time complexities for the computation of $D_{box}$ for both surface representations. \revision{The linear time plots are included in the Supporting Information.} The algorithms for both representation show the trend of growth rate in the order of $\mathcal{O}(C^n)$, with $C > 1$, with the algorithm for the exact representation growing faster than the voxel representation, which is non-ideal for large systems. However, most atomistic objects that Sphractal is designed for typically fall within the range of number of atoms covered by the test cases here. Given this, Figure~\ref{fig:loglogTimeComplexityPlots} indicates that the execution times for both representations are more desirable than algorithms with a linear growth in complexity $\mathcal{O}(n)$. Nonetheless, the practical limit on the number of atoms that could be handled by the algorithms depends on the resources available to the \revision{users}. \par 

\section{Case Study}
\label{sec:caseStudy}

We demonstrate in this section the application of Sphractal in the analysis of 3D fractal dimension of palladium (Pd) nanoparticles, which are promising candidates of catalysts for various industrially relevant reactions~\cite{Lorenzo2012PdCatalysis}. It is recognised that the catalytic activities of metal nanoparticles are impacted by their surface roughness~\cite{Cui2013surfCatalAct,Jeon2021surfORR}, which varies based on the environmental conditions~\cite{Hill2001nanothermodynamics,Baer2011surfChar}. Under electrochemical conditions, metal nanoparticles may undergo leaching of active metals, surface geometric restructuring, and surface compositional restructuring~\cite{Cui2013surfCatalAct,Chee2020geomRestruct}, all of which alter the roughness of their surfaces. However, the dynamic nature of the surface roughness of metal nanoparticles is poorly captured in both experimental and computational studies. We hereby employ Sphractal to compare palladium nanoparticle surfaces quantitatively by computing their $D_{box}$ values. All codes written and data generated here are publicly available at \url{https://github.com/jon-ting/sphractal-methodology-publication}. \par 

\begin{figure}[htbp]  \centering
\sidesubfloat[]{\includegraphics[width=0.45\textwidth]{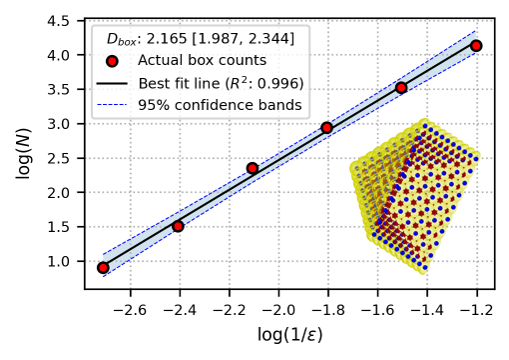}\label{fig:OTS1T000_VX}}
\sidesubfloat[]{\includegraphics[width=0.45\textwidth]{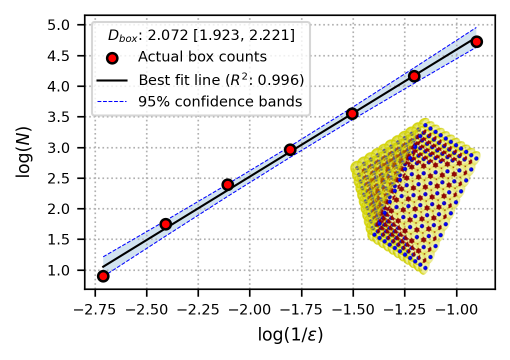}\label{fig:OTS2T000_VX}} \\
\sidesubfloat[]{\includegraphics[width=0.45\textwidth]{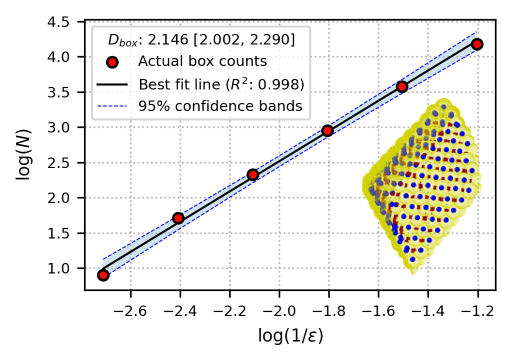}\label{fig:OTS1T323_VX}}
\sidesubfloat[]{\includegraphics[width=0.45\textwidth]{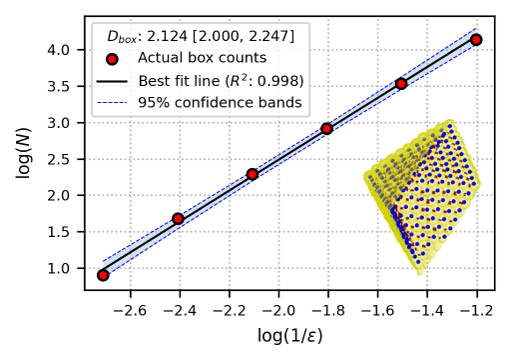}\label{fig:OTS2T323_VX}} \\
\sidesubfloat[]{\includegraphics[width=0.45\textwidth]{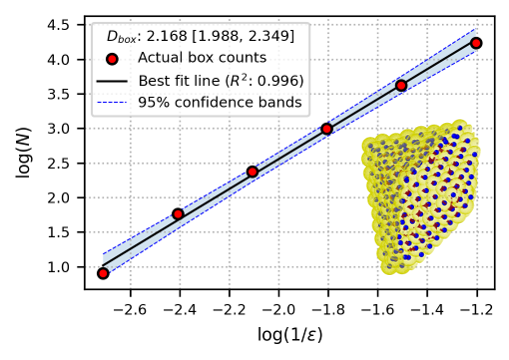}\label{fig:OTS1T523_VX}}
\sidesubfloat[]{\includegraphics[width=0.45\textwidth]{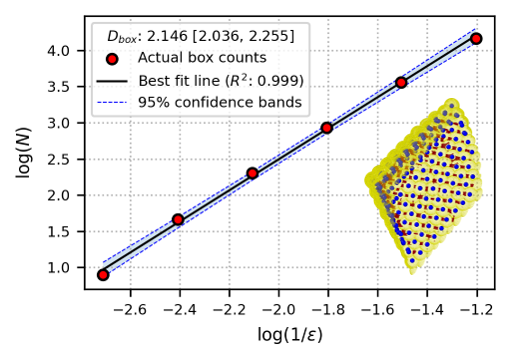}\label{fig:OTS2T523_VX}} \\
\caption{Ordinary least squares best fit lines to box-counting data collected from the voxel representation of the surfaces of octahedral palladium nanoparticles.}
\label{fig:OT_NP_VX_Dbox}
\end{figure}

\begin{figure}[htbp]  \centering
\sidesubfloat[]{\includegraphics[width=0.45\textwidth]{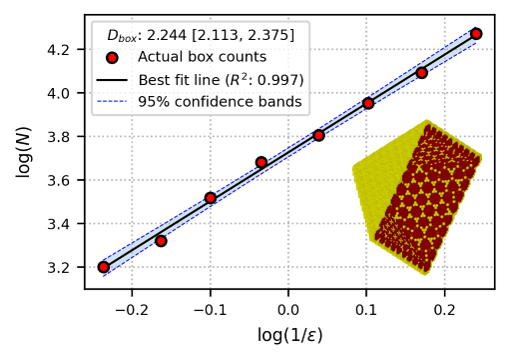}\label{fig:OTS1T000_EX}}
\sidesubfloat[]{\includegraphics[width=0.45\textwidth]{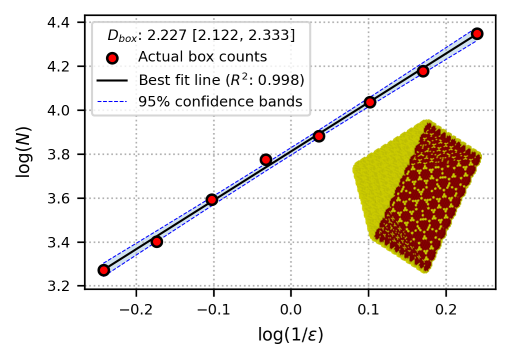}\label{fig:OTS2T000_EX}} \\
\sidesubfloat[]{\includegraphics[width=0.45\textwidth]{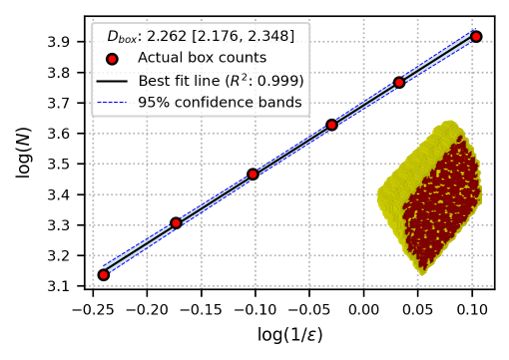}\label{fig:OTS1T323_EX}}
\sidesubfloat[]{\includegraphics[width=0.45\textwidth]{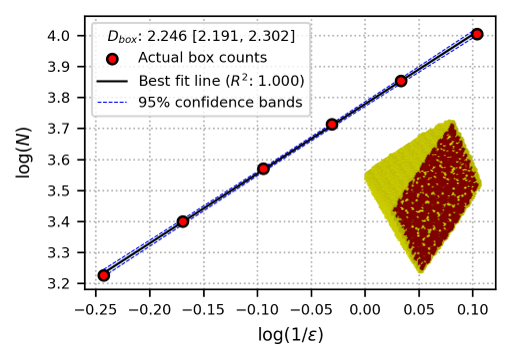}\label{fig:OTS2T323_EX}} \\
\sidesubfloat[]{\includegraphics[width=0.45\textwidth]{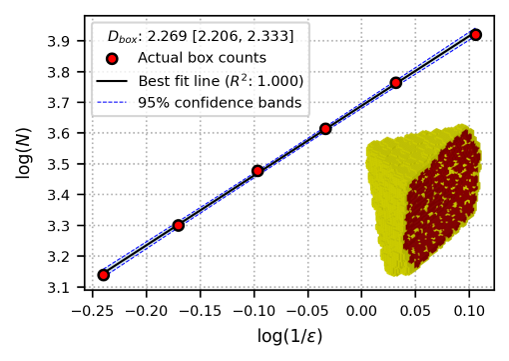}\label{fig:OTS1T523_EX}}
\sidesubfloat[]{\includegraphics[width=0.45\textwidth]{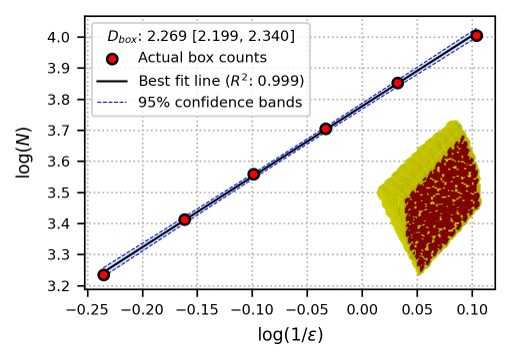}\label{fig:OTS2T523_EX}} \\
\caption{Ordinary least squares best fit lines to box-counting data collected from the exact representation of the surfaces of octahedral palladium nanoparticles.}
\label{fig:OT_NP_EX_Dbox}
\end{figure}

\begin{figure}[htbp]  \centering
\sidesubfloat[]{\includegraphics[width=0.45\textwidth]{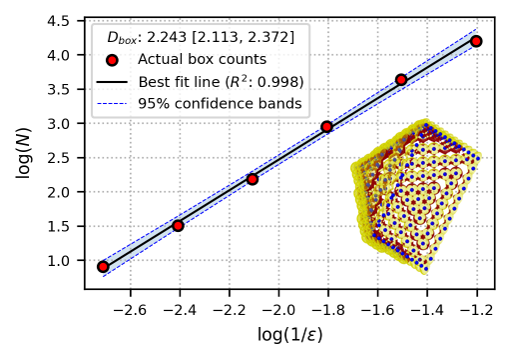}\label{fig:RDS1T000_VX}}
\sidesubfloat[]{\includegraphics[width=0.45\textwidth]{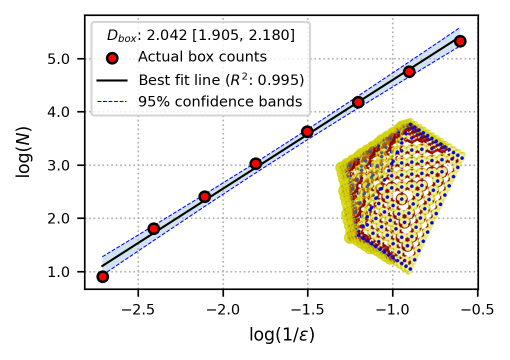}\label{fig:RDS2T000_VX}} \\
\sidesubfloat[]{\includegraphics[width=0.45\textwidth]{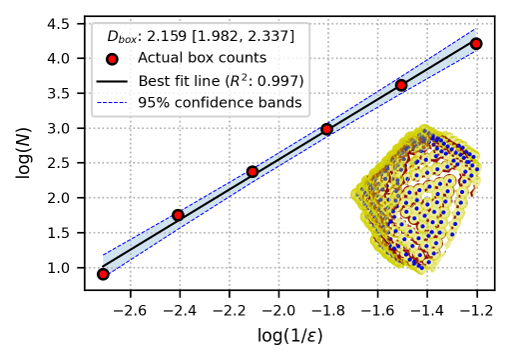}\label{fig:RDS1T323_VX}}
\sidesubfloat[]{\includegraphics[width=0.45\textwidth]{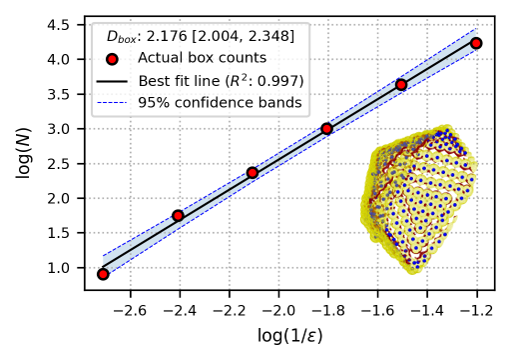}\label{fig:RDS2T323_VX}} \\
\sidesubfloat[]{\includegraphics[width=0.45\textwidth]{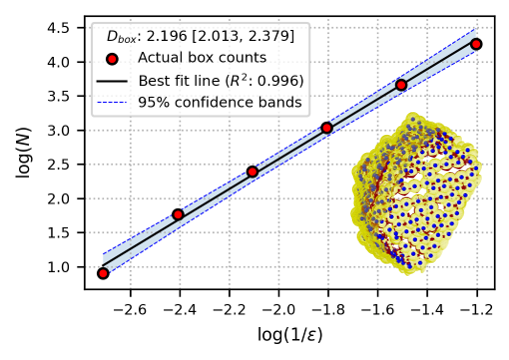}\label{fig:RDS1T523_VX}}
\sidesubfloat[]{\includegraphics[width=0.45\textwidth]{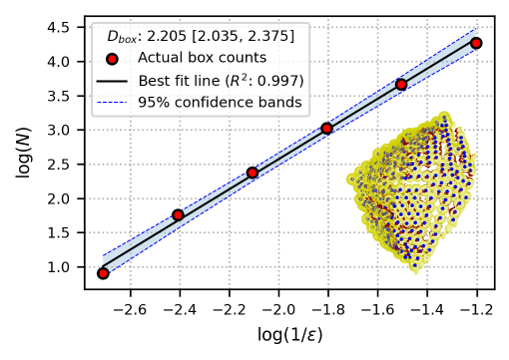}\label{fig:RDS2T523_VX}} \\
\caption{Ordinary least squares best fit lines to box-counting data collected from the voxel representation of the surfaces of rhombic dodecahedral palladium nanoparticles.}
\label{fig:RD_NP_VX_Dbox}
\end{figure}

\begin{figure}[htbp]  \centering
\sidesubfloat[]{\includegraphics[width=0.45\textwidth]{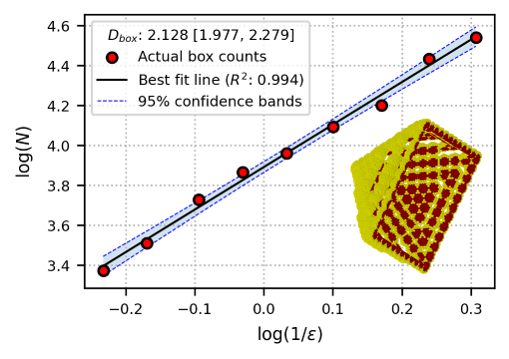}\label{fig:RDS1T000_EX}}
\sidesubfloat[]{\includegraphics[width=0.45\textwidth]{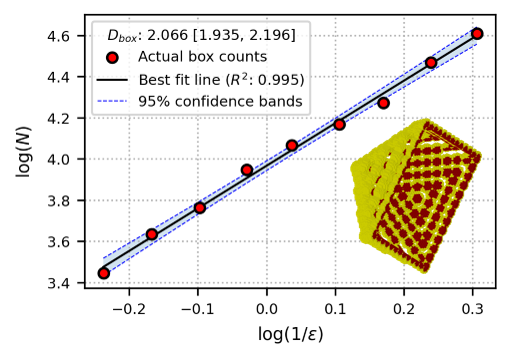}\label{fig:RDS2T000_EX}} \\
\sidesubfloat[]{\includegraphics[width=0.45\textwidth]{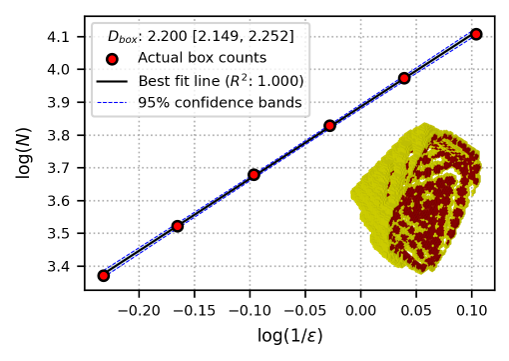}\label{fig:RDS1T323_EX}}
\sidesubfloat[]{\includegraphics[width=0.45\textwidth]{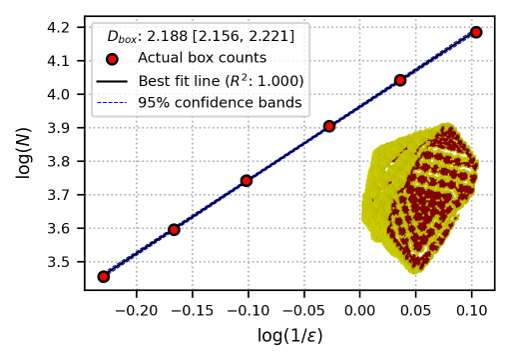}\label{fig:RDS2T323_EX}} \\
\sidesubfloat[]{\includegraphics[width=0.45\textwidth]{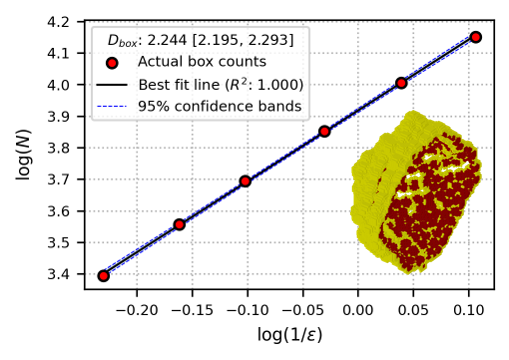}\label{fig:RDS1T523_EX}}
\sidesubfloat[]{\includegraphics[width=0.45\textwidth]{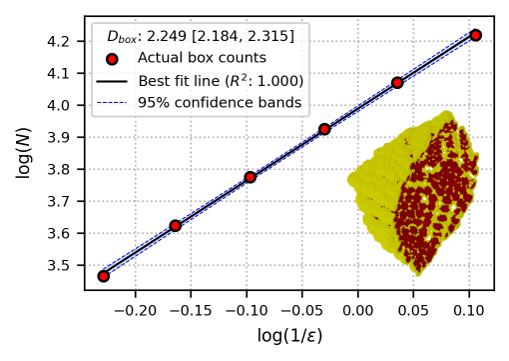}\label{fig:RDS2T523_EX}} \\
\caption{Ordinary least squares best fit lines to box-counting data collected from the exact representation of the surfaces of rhombic dodecahedral palladium nanoparticles.}
\label{fig:RD_NP_EX_Dbox}
\end{figure}

\begin{figure}[htbp]  \centering
\sidesubfloat[]{\includegraphics[width=0.45\textwidth]{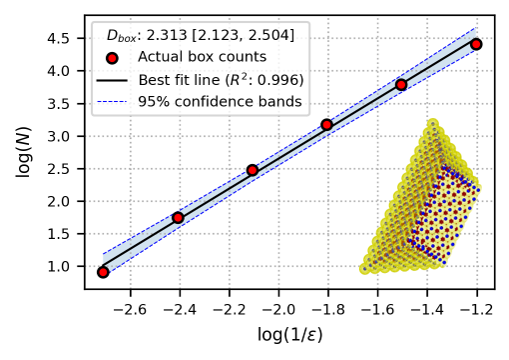}\label{fig:THS1T000_VX}}
\sidesubfloat[]{\includegraphics[width=0.45\textwidth]{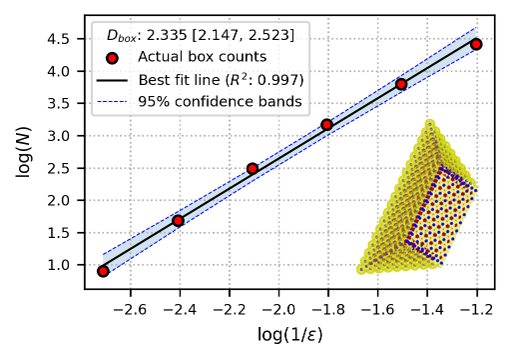}\label{fig:THS2T000_VX}} \\
\sidesubfloat[]{\includegraphics[width=0.45\textwidth]{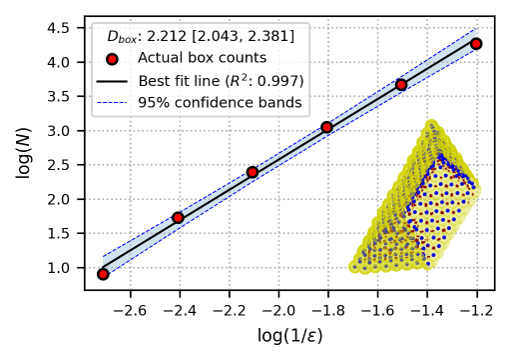}\label{fig:THS1T323_VX}}
\sidesubfloat[]{\includegraphics[width=0.45\textwidth]{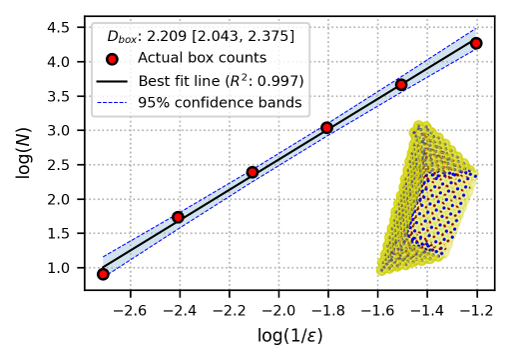}\label{fig:THS2T323_VX}} \\
\sidesubfloat[]{\includegraphics[width=0.45\textwidth]{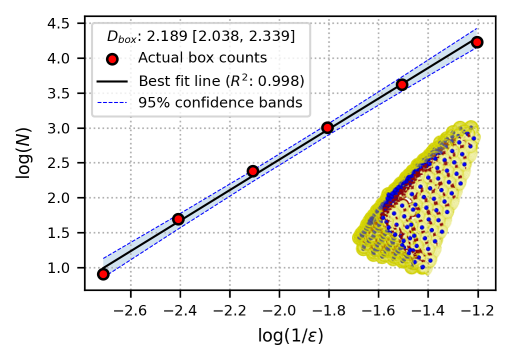}\label{fig:THS1T523_VX}}
\sidesubfloat[]{\includegraphics[width=0.45\textwidth]{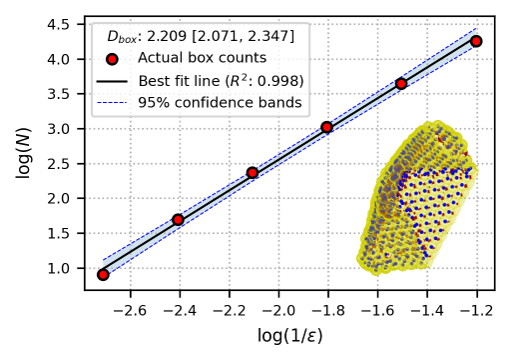}\label{fig:THS2T523_VX}} \\
\caption{Ordinary least squares best fit lines to box-counting data collected from the voxel representation of the surfaces of tetrahedral palladium nanoparticles.}
\label{fig:TH_NP_VX_Dbox}
\end{figure}

\begin{figure}[htbp]  \centering
\sidesubfloat[]{\includegraphics[width=0.45\textwidth]{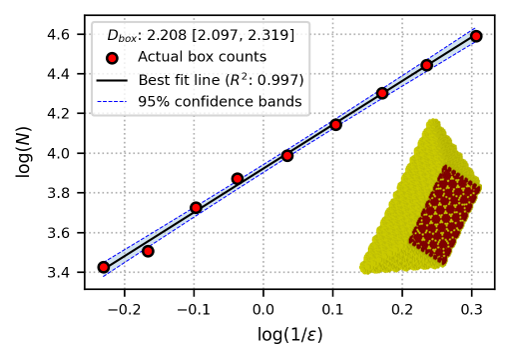}\label{fig:THS1T000_EX}}
\sidesubfloat[]{\includegraphics[width=0.45\textwidth]{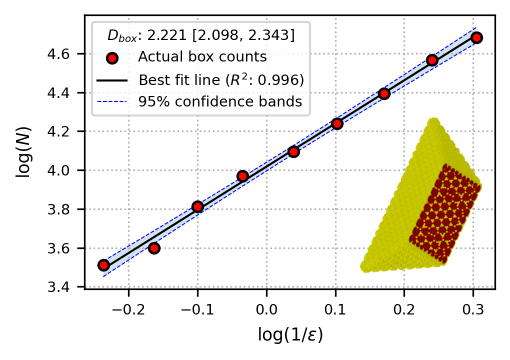}\label{fig:THS2T000_EX}} \\
\sidesubfloat[]{\includegraphics[width=0.45\textwidth]{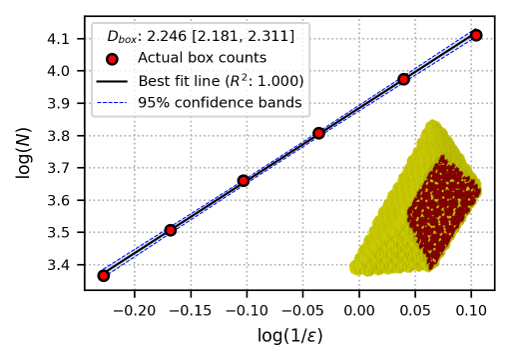}\label{fig:THS1T323_EX}}
\sidesubfloat[]{\includegraphics[width=0.45\textwidth]{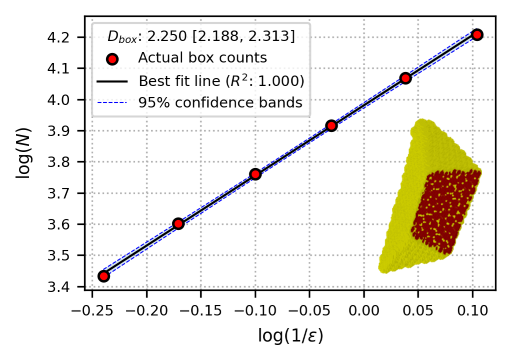}\label{fig:THS2T323_EX}} \\
\sidesubfloat[]{\includegraphics[width=0.45\textwidth]{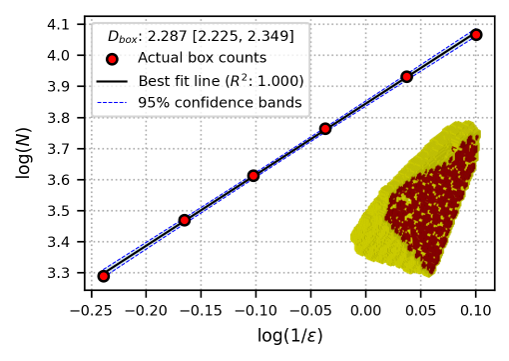}\label{fig:THS1T523_EX}}
\sidesubfloat[]{\includegraphics[width=0.45\textwidth]{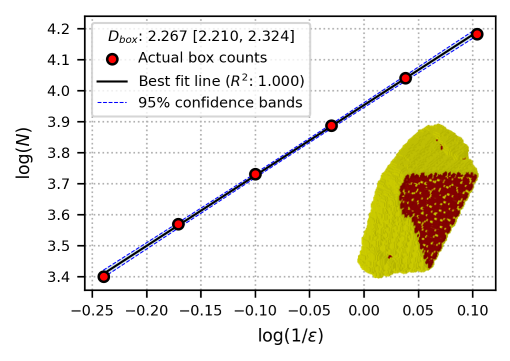}\label{fig:THS2T523_EX}} \\
\caption{Ordinary least squares best fit lines to box-counting data collected from the exact representation of the surfaces of tetrahedral palladium nanoparticles.}
\label{fig:TH_NP_EX_Dbox}
\end{figure}

The test cases used here were obtained from a publicly available dataset, originally collected to investigate the impact of polydispersivity on the properties of Pd nanoparticle electrocatalyst ensembles~\cite{Barnard2019PdData}. Three degrees of freedom were varied among the test cases chosen for comparisons, namely shape, size, and temperature, as illustrated in Table~\ref{tab:Dbox_PdNPs}. The $D_{box}$ values were computed using the default parameter values. The ordinary least squares fitted to the box-counting data collected for all of the test cases are shown in Figures~\ref{fig:OT_NP_VX_Dbox} to \ref{fig:TH_NP_EX_Dbox}. \par

\begin{longtable}{ccccccc}
\hline
\multicolumn{1}{c}{Shape} & \multicolumn{1}{c}{Size} &\multicolumn{1}{c}{Temp (K)} & \multicolumn{2}{c}{Voxel Representation} & \multicolumn{2}{c}{Exact Representation} \\
\cline{4-7}
\multicolumn{1}{c}{} & \multicolumn{1}{c}{} & \multicolumn{1}{c}{} & $\boldsymbol{D_{b_{V}}}$ [95\% CI] & $R^2$ & $\boldsymbol{D_{b_{E}}}$ [95\% CI] & $R^2$ \\ \hline
OT & Small & 0 & \textbf{2.165} [1.987, 2.344] & 0.996 & \textbf{2.244} [2.113, 2.375] & 0.997 \\
OT & Small & 323 & \textbf{2.146} [2.002, 2.290] & 0.998 & \textbf{2.262} [2.176, 2.348] & 0.999 \\
OT & Small & 523 & \textbf{2.169} [1.988, 2.349] & 0.996 & \textbf{2.269} [2.206, 2.333] & 1.000 \\
\cline{2-7}
OT & Large & 0 & \textbf{2.072} [1.923, 2.221] & 0.996 & \textbf{2.227} [2.122, 2.333] & 0.998 \\
OT & Large & 323 & \textbf{2.124} [2.000, 2.247] & 0.998 & \textbf{2.246} [2.191, 2.302] & 1.000 \\
OT & Large & 523 & \textbf{2.146} [2.036, 2.255] & 0.999 & \textbf{2.269} [2.199, 2.340] & 0.999 \\
\hline
RD & Small & 0 & \textbf{2.243} [2.113, 2.372] & 0.998 & \textbf{2.128} [1.977, 2.279] & 0.994 \\
RD & Small & 323 & \textbf{2.159} [1.982, 2.337] & 0.997 & \textbf{2.200} [2.149, 2.252] & 1.000 \\
RD & Small & 523 & \textbf{2.196} [2.013, 2.379] & 0.996 & \textbf{2.244} [2.195, 2.293] & 1.000 \\
\cline{2-7}
RD & Large & 0 & \textbf{2.042} [1.905, 2.180] & 0.995 & \textbf{2.066} [1.935, 2.196] & 0.995 \\
RD & Large & 323 & \textbf{2.176} [2.004, 2.348] & 0.997 & \textbf{2.188} [2.156, 2.221] & 1.000 \\
RD & Large & 523 & \textbf{2.205} [2.035, 2.375] & 0.997 & \textbf{2.249} [2.184, 2.315] & 1.000 \\
\hline
TH & Small & 0 & \textbf{2.313} [2.123, 2.504] & 0.996 & \textbf{2.208} [2.097, 2.319] & 0.997 \\
TH & Small & 323 & \textbf{2.212} [2.043, 2.381] & 0.997 & \textbf{2.246} [2.181, 2.311] & 1.000 \\
TH & Small & 523 & \textbf{2.189} [2.038, 2.339] & 0.998 & \textbf{2.287} [2.225, 2.349] & 1.000 \\
\cline{2-7}
TH & Large & 0 & \textbf{2.335} [2.147, 2.523] & 0.997 & \textbf{2.221} [2.098, 2.343] & 0.996 \\
TH & Large & 323 & \textbf{2.209} [2.043, 2.375] & 0.997 & \textbf{2.250} [2.188, 2.313] & 1.000 \\
TH & Large & 523 & \textbf{2.209} [2.071, 2.347] & 0.998 & \textbf{2.267} [2.210, 2.324] & 1.000 \\
\hline
\caption{Box-counting dimensions of palladium nanoparticles with different shapes, namely octahedron (OT), rhombic dodecahedron (RD), and tetrahedron (TH). The confidence intervals are computed based on Student's $t$-distribution using functionalities offered by the \texttt{statsmodels} package.}
\label{tab:Dbox_PdNPs}
\end{longtable}

In accordance with the literature~\cite{Laaksonen2016fracSurf}, all $D_{box}$ values obtained are bounded within $[2, 3]$. The coefficient of determination $R^2$ is also consistently above 0.99, indicating that the slopes conform well to linearity, providing $D_{box}$ estimation with low uncertainty. The magnitude of the box-counting dimension computed from the exact surface representation, $D_{b_{E}}$, is found to be directly proportional to the temperature of the nanoparticle conformation. As one would expect, nanoparticle conformations at higher temperatures will be more disordered due to a higher degree of irregularities on the surface, and therefore tend to have higher $D_{b_{E}}$ values than their counterpart ordered conformations. \par 

In contrast, the box-counting dimensions computed from the voxel representation, $D_{b_{V}}$, do not exhibit similar capability in detecting the trend. However, it complements the exact representation by capturing the surface patterns over a larger scale (wider range of box lengths) with cheaper computational costs. Researchers have arrived at a consensus that an object is said to exhibit fractal properties only if its log-log plot shows linearity over the span of two or more decades (a decade being a ratio of 10)~\cite{Kenkel2013FDdecade}. Examination of the log-log plots prior to box count data removal (for the determination of the range of box lengths that yields the highest $R^2$) confirmed that all of the test cases could be considered as fractals, exhibiting linearity over more than two decades. The log-log plots without point removal have been included in the Supporting Information. \par 

Given these observations, $D_{b_{V}}$ could be computed in a high-throughput study to assess whether a given nanoparticle surface is likely to exhibit fractal behaviour prior to conducting the more expensive computation of $D_{b_{E}}$. \par

\section{Conclusion}
This article presents two approaches to estimate the fractal dimension of the surface of three-dimensional objects composed of spheres, by computing the box-counting dimensions of their representations as either voxelised point clouds or mathematically precise objects. The earlier representation is cheaper computationally and captures the surface complexity at a larger scale, allowing the \revision{users} to assess whether a given object exhibits fractal behaviour. The latter is able to provide insights into the complexity of the surface at a specific range of scale that is of interest to the \revision{users}, enabling quantitative comparisons among similar objects. Both of the algorithms are entirely deterministic. The methodology was implemented and published as Sphractal, a freely available platform agnostic Python package. \par 

We validated the implementation of the method by comparing the results to the theoretical fractal dimension of a perfect sphere. The usefulness of Sphractal in capturing atomistic surface complexities at the nanoscale was demonstrated by applying it to a simulated palladium nanoparticles data set. 

By providing a quantitative means to measure the details of the protrusions and indentations on the surfaces of individual atomistic objects as obtained from computational studies, it is expected that Sphractal could contribute to the understanding of the impact of the surface of atomistic objects such as metal nanoparticles towards their properties. The package is designed to be generalisable to other systems where the components are represented as spheres, such as biomolecular systems. However, care is needed to interpret the box-counting dimensions computed. \par

The open-source nature of Sphractal allows for broad collaborations within the scientific community. The package will continue to be developed and have features added over time, particularly those related to its scalability. These include parallelisation of the algorithms over multiple compute nodes and further parallelisation of other steps such as the identification of surface atoms. As the resulting box-counting dimension is strongly dependent on the correct identification of surface atoms, a future task involves the incorporation of alternative surface atom identification algorithms~\cite{Hinuma2020findSurf,Wang2005surfConeAlg,Kim2011findSurf,Sega2018findSurf}, which might better suit other types of atomistic objects. \par

\medskip
\textbf{Supporting Information} \par
Supporting Information is available from the Wiley Online Library or from the author.

\medskip
\textbf{Acknowledgements} \par
Computational resources for this project were supplied by the National Computing Infrastructure (NCI) [grant numbers p00 and q27]. Jonathan Y. C. Ting is grateful for the support from Australian National University via the University Research Scholarship. This work was completed in part at the NCI-NVIDIA Open Hackathon 2023, part of the Open Hackathons program. The authors would like to acknowledge OpenACC-Standard.org for their support.

\medskip
\textbf{Declaration of Competing Interest} \par
The authors declare no conflict of interest.

\medskip
\textbf{Data Availability} \par
The source codes of Sphractal are publicly available at \url{https://github.com/jon-ting/sphractal} and \url{https://github.com/jon-ting/fastbc}. Other data and written codes that support the findings of this study are openly available at \url{https://github.com/jon-ting/sphractal-methodology-publication}.

\medskip
\bibliographystyle{MSP}
\bibliography{ref}

\begin{figure}
\textbf{Table of Contents}\\
\medskip
  \includegraphics{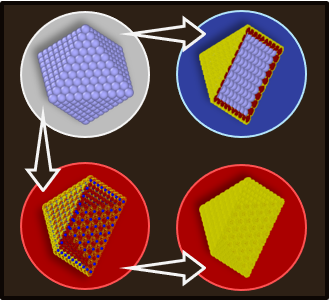}
  \medskip
  \caption*{The fractal dimension of a surface allows its degree of roughness to be characterised quantitatively. The three-dimensional surface computed from precise atomic coordinates can be represented as either a voxelised point cloud or a mathematically exact surface. Sphractal is a Python package that estimates the fractal dimensions of such surfaces by computing their box-counting dimensions.}
\end{figure}

\end{document}


\title{ \bf SUPPORTING INFORMATION \\ Sphractal: Estimating the Fractal Dimension of Surfaces Computed from Precise Atomic Coordinates via Box-Counting Algorithm}
\author[1]{Jonathan Yik Chang Ting \footnote{Corresponding author: jonathan.ting@anu.edu.au}}
\author[2]{Andrew Thomas Agars Wood}
\author[1]{Amanda Susan Barnard}
\affil[1]{School of Computing, Australian National University, Acton 2601, Australia}
\affil[2]{Research School of Finance, Actuarial Studies \& Statistics, Australian National University, Acton 2601, Australia}
\date{}
\setcounter{Maxaffil}{0}
\renewcommand\Affilfont{\itshape\small}

\maketitle

Figure~\ref{fig:boxPlots} shows the box plots of the run time distributions for all computational complexity and scalability tests conducted, where the outliers, means, and medians were marked.

\begin{figure}[htbp]  \centering
\includegraphics[width=0.45\textwidth]{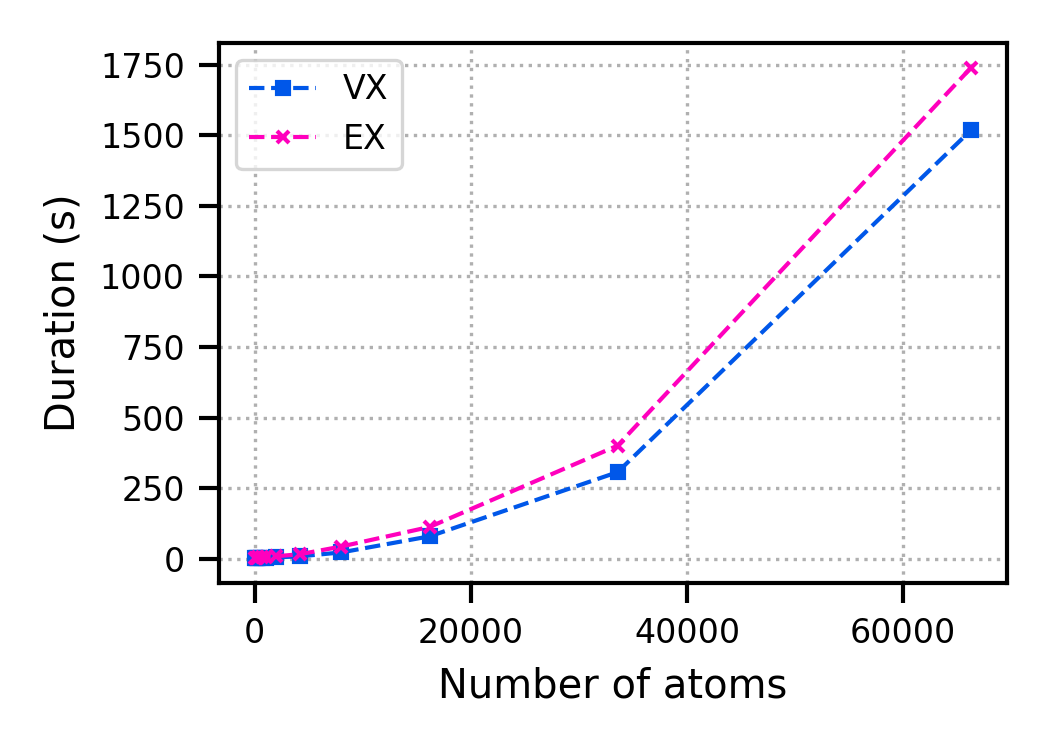}
\includegraphics[width=0.45\textwidth]{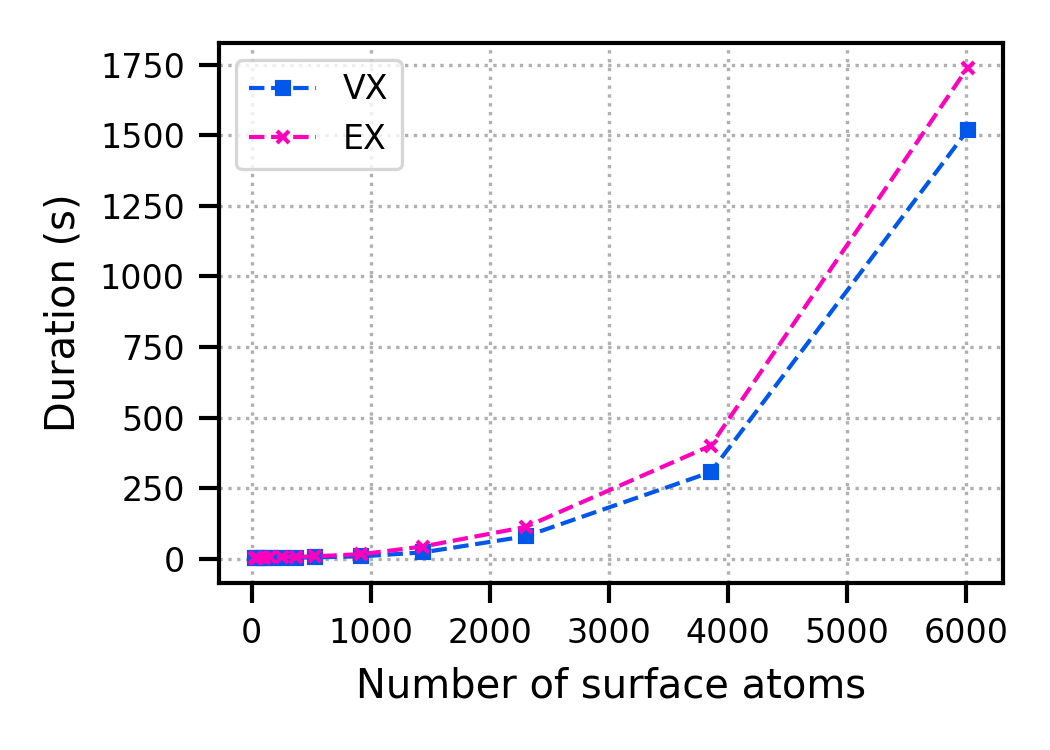} \\
\caption{Linear time plots for box-counting of the voxelised point cloud and mathematically exact surface representations, as a function of (a) total number of atoms and (b) number of surface atoms.}
\label{fig:linearTimeComplexityPlots}
\end{figure}

Figure~\ref{fig:linearTimeComplexityPlots} plots the time complexity in linear time, for comparison with the logarithmic time plots in the main text.

\begin{figure}[htbp]  \centering
\sidesubfloat[]{\includegraphics[width=0.45\textwidth]{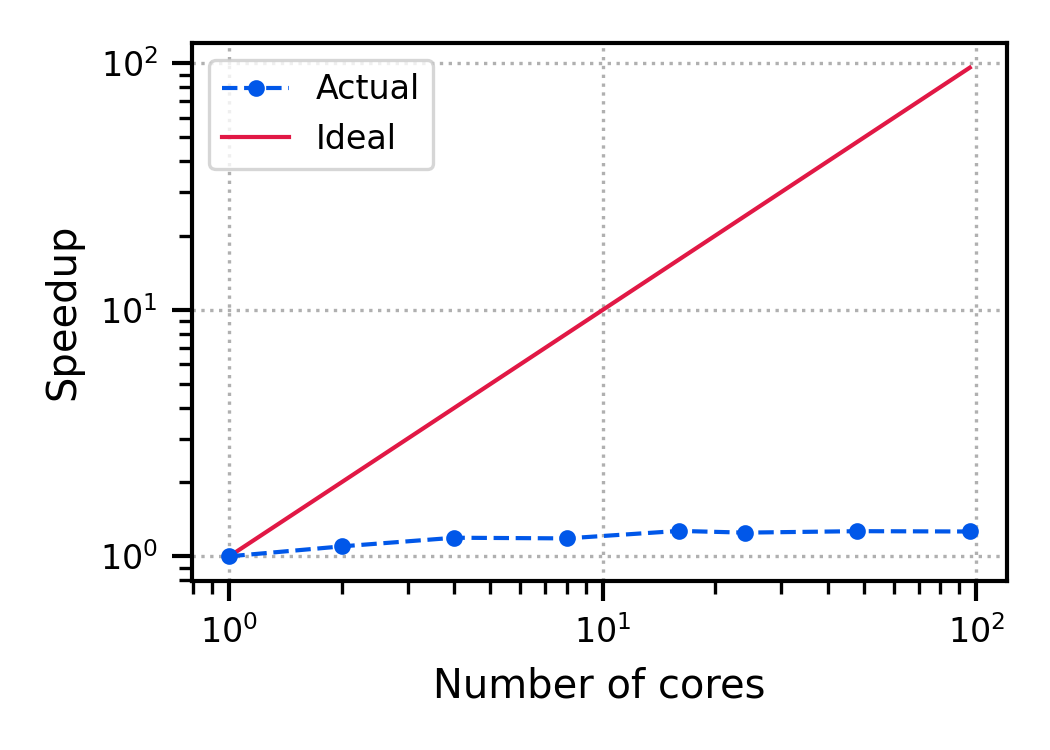}\label{fig:SS_VX_Speedups}}
\sidesubfloat[]{\includegraphics[width=0.45\textwidth]{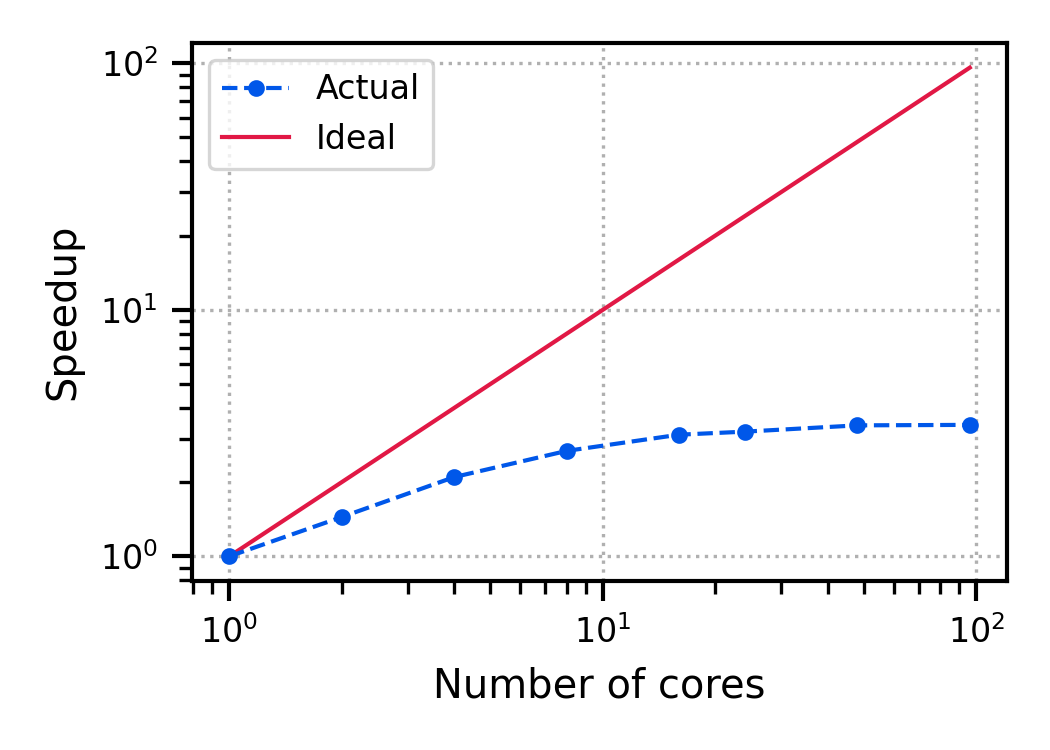}\label{fig:SS_EX_Speedups}} \\
\sidesubfloat[]{\includegraphics[width=0.45\textwidth]{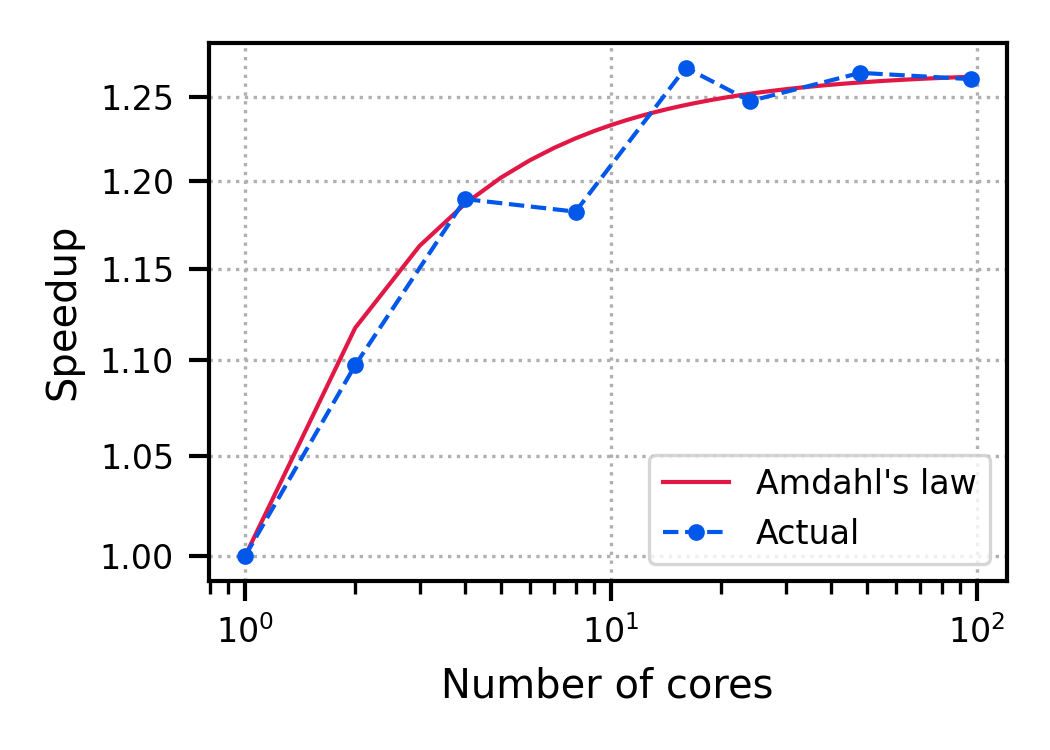}\label{fig:SS_VX_Speedup}}
\sidesubfloat[]{\includegraphics[width=0.45\textwidth]{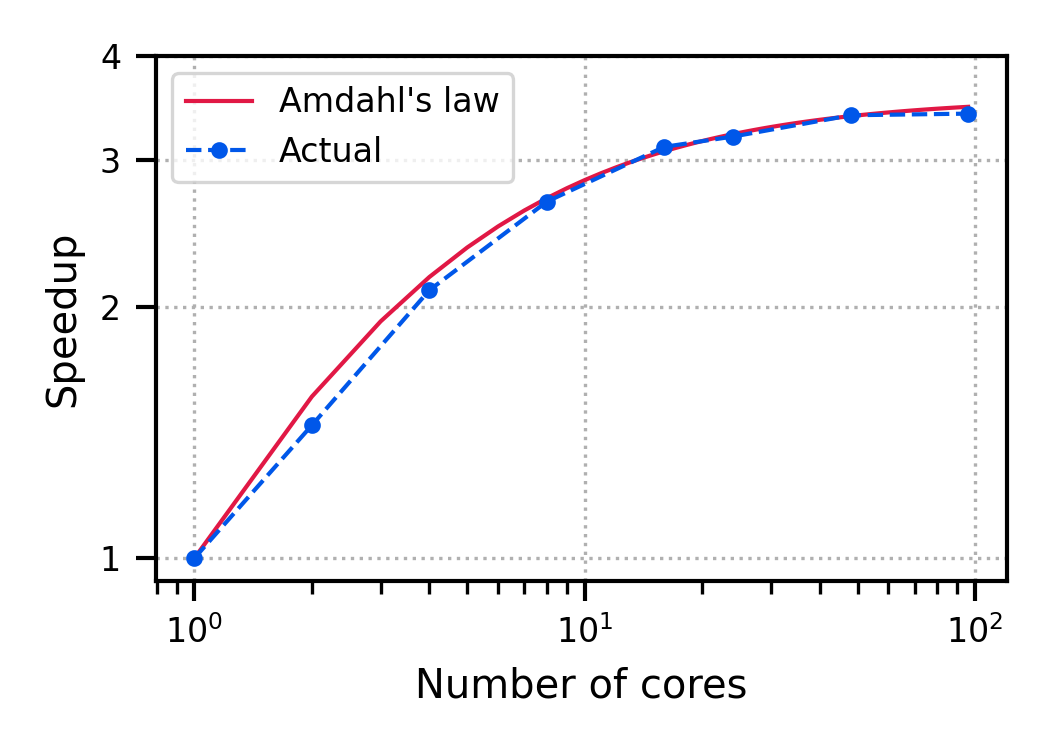}\label{fig:SS_EX_Speedup}} \\
\caption{Strong scaling plots showing (a \& b) comparison between ideal (linear speedup) and actual speedups, and (c \& d) comparison between manually fitted Amdahl's law curves and actual speedups of the box-counting algorithms for the voxelised point cloud (left) and mathematically exact (right) surface representations, respectively.}
\label{fig:strongScalingPlots}
\end{figure}

Figure~\ref{fig:strongScalingPlots} shows the comparisons between the actual strong scaling speedups of the box-counting algorithms for both surface representations with the ideal linear speedups and the manual fit of Amdahl's law curves, which describe the theoretical maximum speedup (Equation~\ref{eq:Amdahl}). The strong scaling speedups were calculated by keeping a constant problem size while increasing the number of cores, which results in a reduced workload per processor. A relatively large nanoparticle (spherical palladium nanoparticle with a diameter of 100 \AA, which has an atom count of 35529) is chosen as a test case for strong scaling to ensure full occupation of each node when a large number of cores is used.  \par 

\begin{equation}
\label{eq:Amdahl}
S_{max} = \frac{1}{(1-p) + p/s}
\end{equation}
where: 

\hspace{24pt} $S_{max}$ = theoretical maximum speedup of the execution of the whole task
  
\hspace{24pt} $p$ = proportion of run time that the parallelisable parts occupied 
  
\hspace{24pt} $s$ = speedup of the parallelisable part that benefits from parallelisation (assumed to be linearly correlated with the number of cores used) 

The greater speedup magnitude for the exact representation is due to its larger proportion of parallelisable processes than the voxel representation. This is because the whole box-counting procedure for the former is embarrassingly parallel, while the parallelisation for the latter only involves the point cloud generation step. Based on the speedup upper bounds indicated by Figure~\ref{fig:strongScalingPlots}, the proportion, $p$, of the run time that the parallelisable parts occupied are estimated to be about 0.21 (maximum 1.3 times speedup) and 0.72 (maximum 3.6 times speedup) for the voxel and exact surface representations, respectively. The maximum speedup factor is achieved when more cores are used. However, both of the speedup curves have also begin to plateau as the current version of Sphractal could only be parallelised over one compute node due to the lack of support for parallelisation over distributed system in Python. \par

\begin{figure}[htbp]  \centering
\sidesubfloat[]{\includegraphics[width=0.45\textwidth]{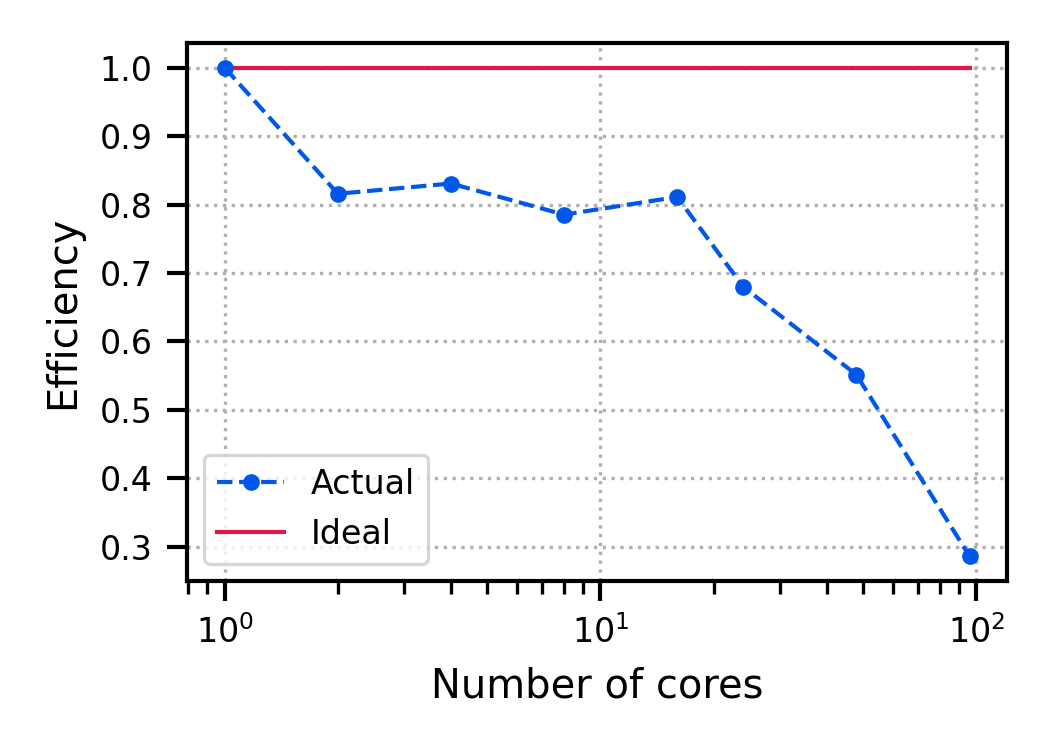}\label{fig:WS_VX_Efficiency}}
\sidesubfloat[]{\includegraphics[width=0.45\textwidth]{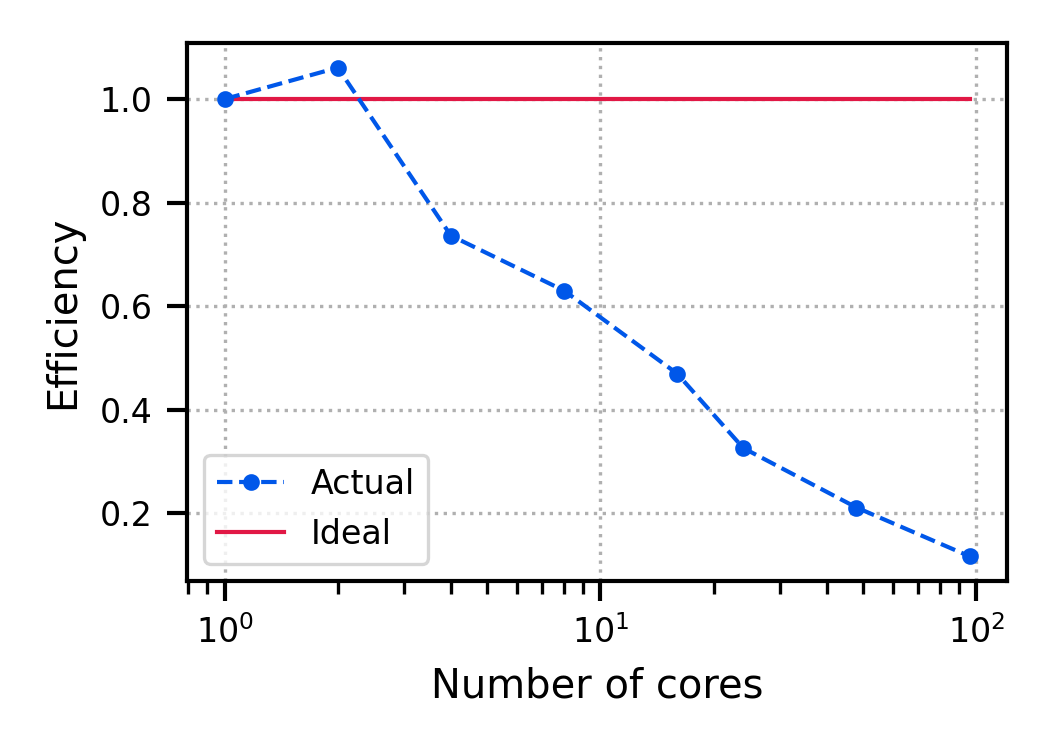}\label{fig:WS_EX_Efficiency}} \\
\caption{Weak scaling plots showing efficiencies of the box-counting algorithms for the (a) voxelised point cloud and (b) mathematically exact surface representations.}
\label{fig:weakScalingPlots}
\end{figure}

Figure~\ref{fig:weakScalingPlots} illustrates the weak scaling efficiencies of the box-counting algorithms for both surface representations. To calculate the weak scaling efficiencies, a constant workload per processor is attempted to be maintained by increasing the problem size and number of cores simultaneously. This is challenging to achieve perfectly in our case as it is difficult to generate simulated (including perfectly spherical) nanoparticles with specific number of atoms. The test cases were therefore chosen among spherical palladium nanoparticles with the closest number of atoms to the ideal setup, where the number of atoms and number of cores increase exponentially such that each processor handles a constant number of atoms for each test case. The diameter of the resulting test case nanoparticles ranges from 10 \AA\ (43 atoms) to 49 \AA\ (4189 atoms). \par

The decline in parallelisation efficiency is greater for the exact representation compared to the voxel representation. This is expected as the overhead for parallelisation would be proportional to the proportion of run time that the parallelisable parts occupied $p$ (estimated to be 0.21 and 0.72 for the former and latter, respectively). The cusp with an apparent efficiency of greater than $1.0$ when two cores are used is likely due to the lower percentage of surface atom counts in the second test case. This is because the proportion of surface atoms, which decides the bulk of the work of each processor, varies with the size of nanoparticle. Given the formulae for the surface area of a sphere ($SA = 4 \pi r^2$) and the volume of a sphere ($V = 4 / 3 \pi r^3$), the ratio of surface area to volume for a sphere ($SA:V = 3 / r$) declines rapidly as the size of our test cases increases, causing the actual workload for each core during the parallelised steps to reduce with increasing nanoparticle size. \par 

\begin{figure}[htbp]  \centering
\includegraphics[width=0.45\textwidth]{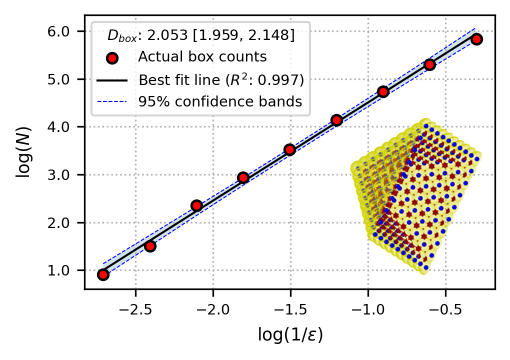}
\includegraphics[width=0.45\textwidth]{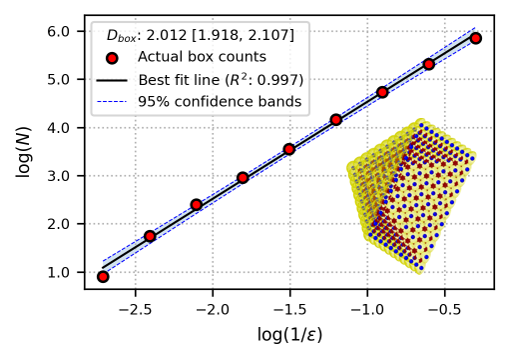} \\
\includegraphics[width=0.45\textwidth]{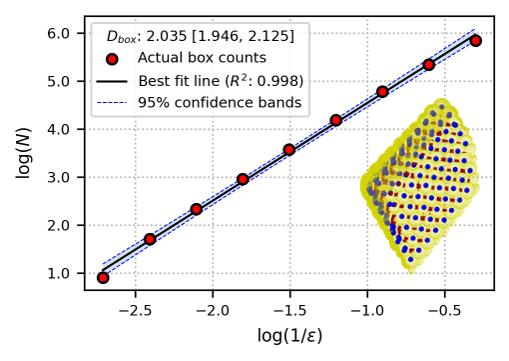}
\includegraphics[width=0.45\textwidth]{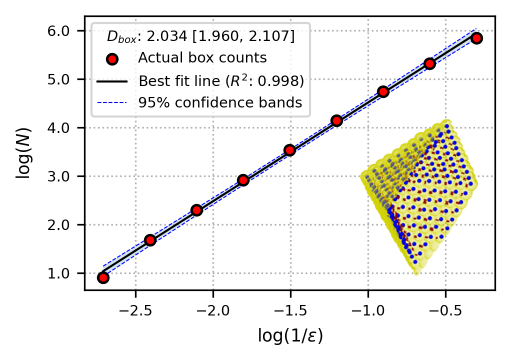} \\
\includegraphics[width=0.45\textwidth]{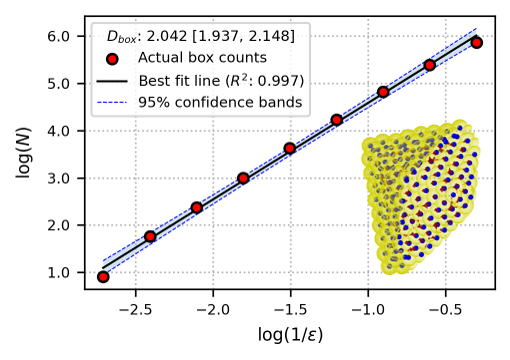}
\includegraphics[width=0.45\textwidth]{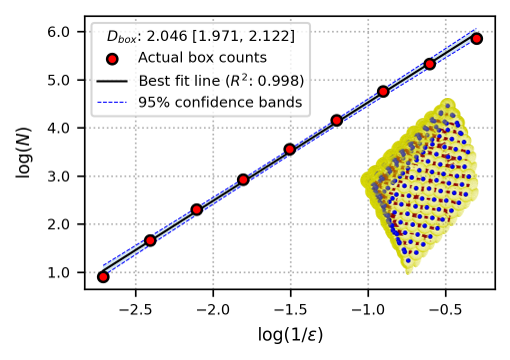} \\
\caption{Full range log-log plots of the linear regression on the box-counting data collected from the voxel representation of octahedral palladium nanoparticle surfaces.}
\label{fig:OT_VX_loglogFullRangePlots}
\end{figure}

\begin{figure}[htbp]  \centering
\includegraphics[width=0.45\textwidth]{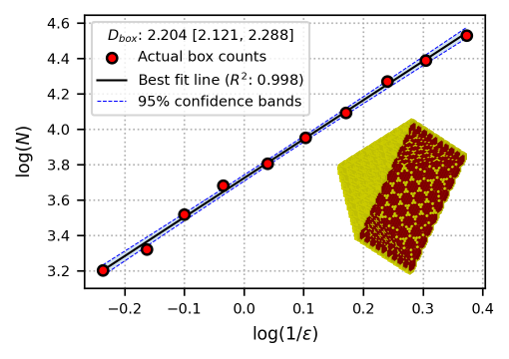}
\includegraphics[width=0.45\textwidth]{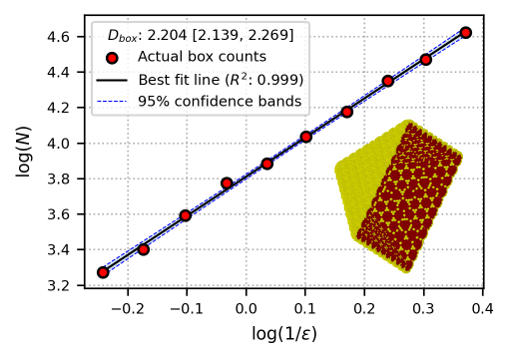} \\
\includegraphics[width=0.45\textwidth]{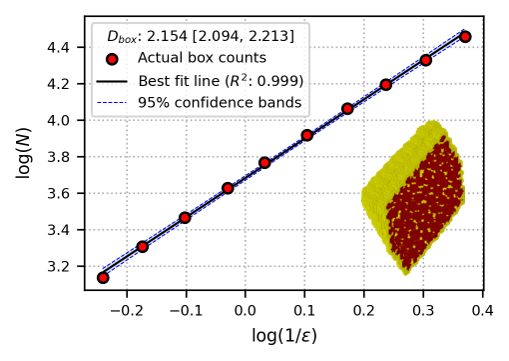}
\includegraphics[width=0.45\textwidth]{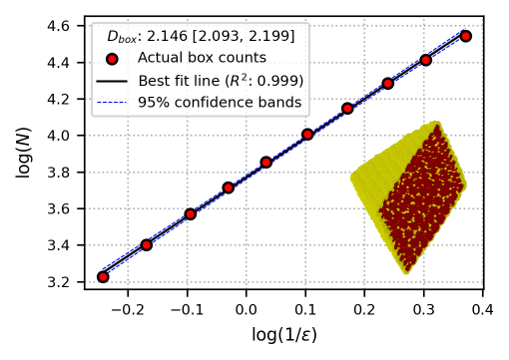} \\
\includegraphics[width=0.45\textwidth]{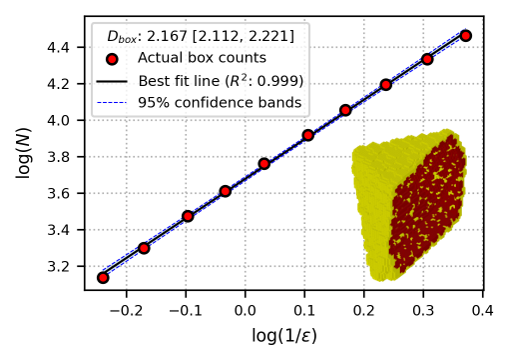}
\includegraphics[width=0.45\textwidth]{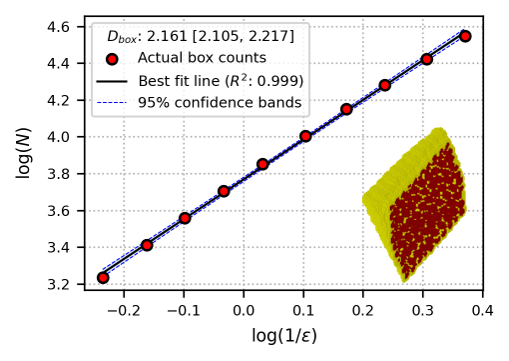} \\
\caption{Full range log-log plots of the linear regression on the box-counting data collected from the exact representation of octahedral palladium nanoparticle surfaces.}
\label{fig:OT_EX_loglogFullRangePlots}
\end{figure}

\begin{figure}[htbp]  \centering
\includegraphics[width=0.45\textwidth]{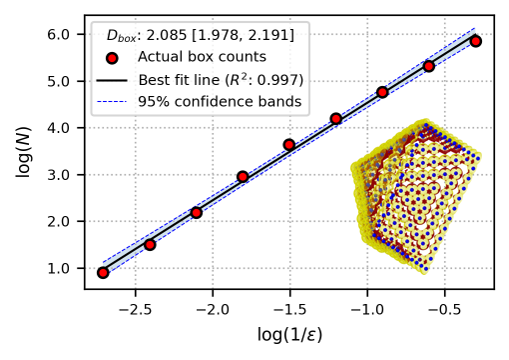}
\includegraphics[width=0.45\textwidth]{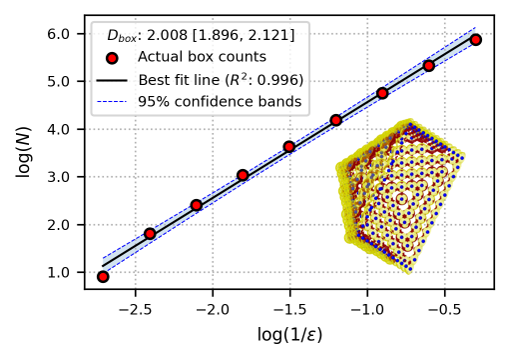} \\
\includegraphics[width=0.45\textwidth]{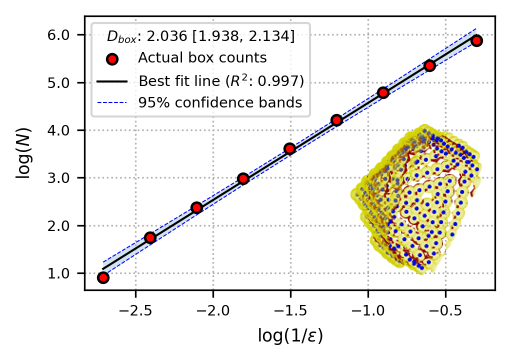}
\includegraphics[width=0.45\textwidth]{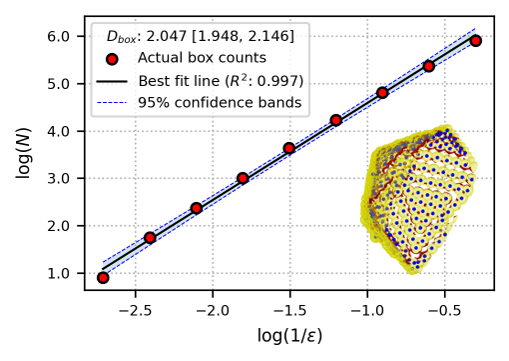} \\
\includegraphics[width=0.45\textwidth]{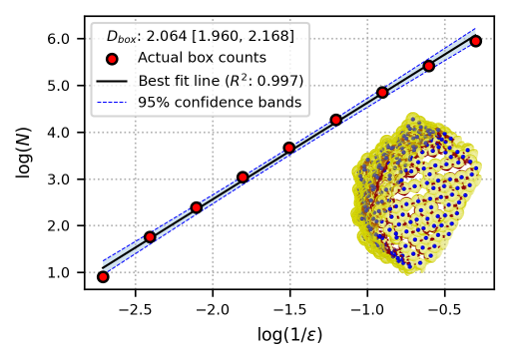}
\includegraphics[width=0.45\textwidth]{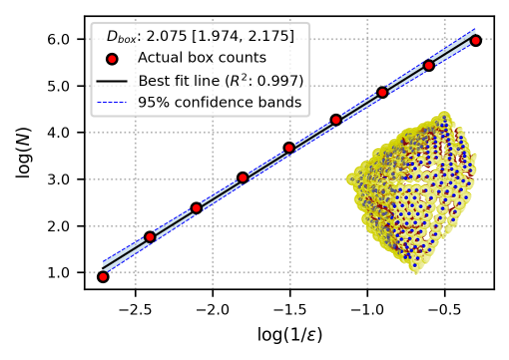} \\
\caption{Full range log-log plots of the linear regression on the box-counting data collected from the voxel representation of rhombic dodecahedral palladium nanoparticle surfaces.}
\label{fig:RD_VX_loglogFullRangePlots}
\end{figure}

\begin{figure}[htbp]  \centering
\includegraphics[width=0.45\textwidth]{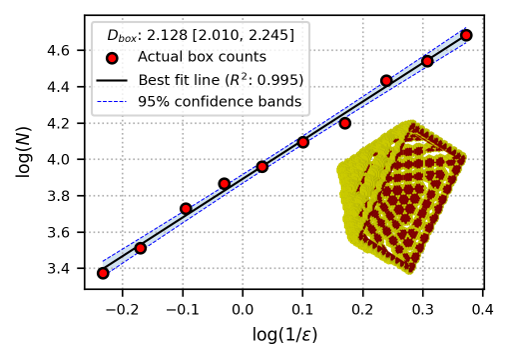}
\includegraphics[width=0.45\textwidth]{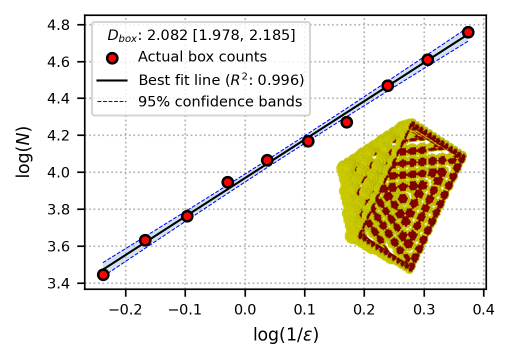} \\
\includegraphics[width=0.45\textwidth]{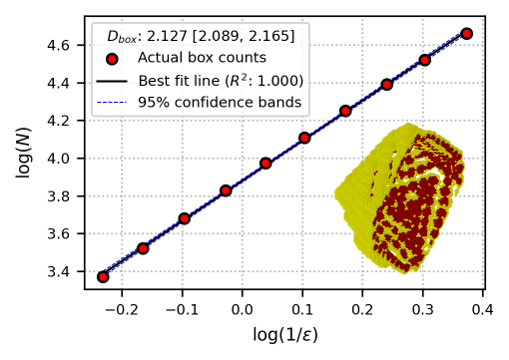}
\includegraphics[width=0.45\textwidth]{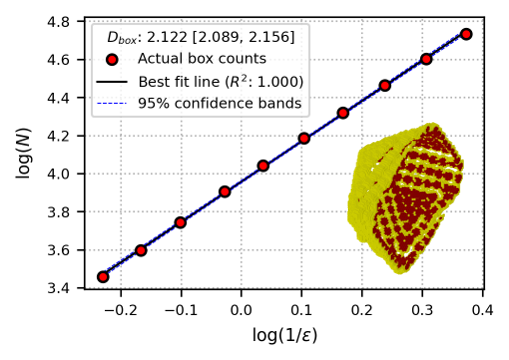} \\
\includegraphics[width=0.45\textwidth]{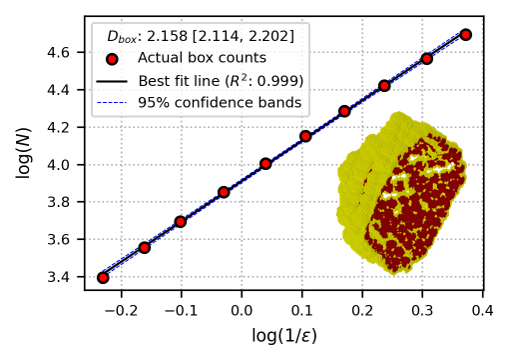}
\includegraphics[width=0.45\textwidth]{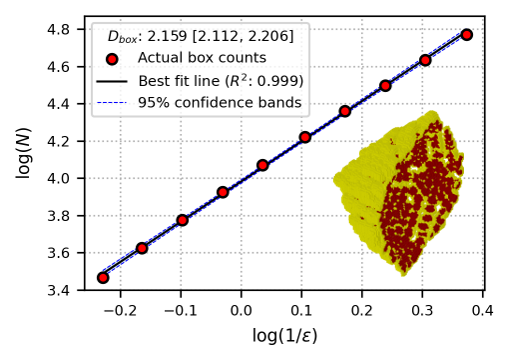} \\
\caption{Full range log-log plots of the linear regression on the box-counting data collected from the exact representation of rhombic dodecahedral palladium nanoparticle surfaces.}
\label{fig:RD_EX_loglogFullRangePlots}
\end{figure}

\begin{figure}[htbp]  \centering
\includegraphics[width=0.45\textwidth]{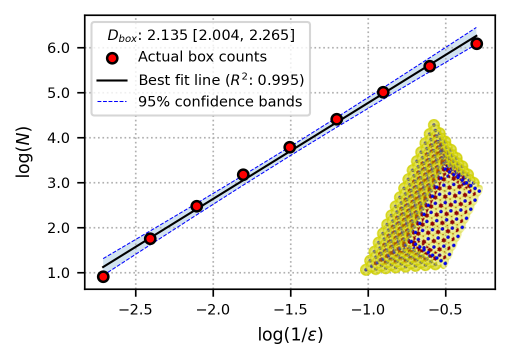}
\includegraphics[width=0.45\textwidth]{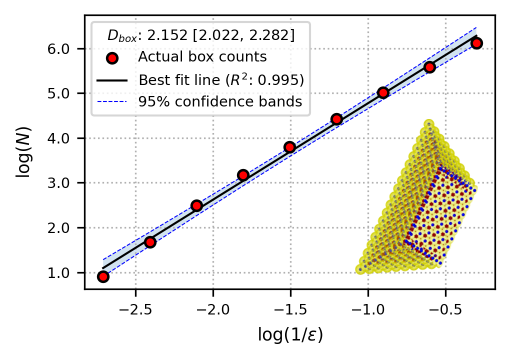} \\
\includegraphics[width=0.45\textwidth]{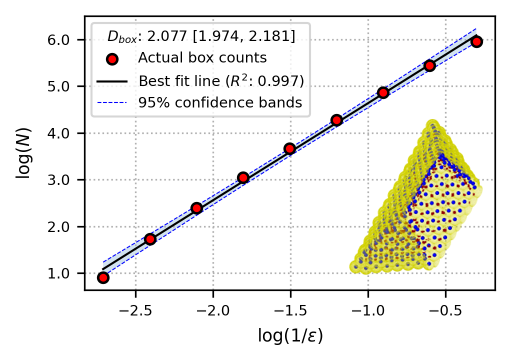}
\includegraphics[width=0.45\textwidth]{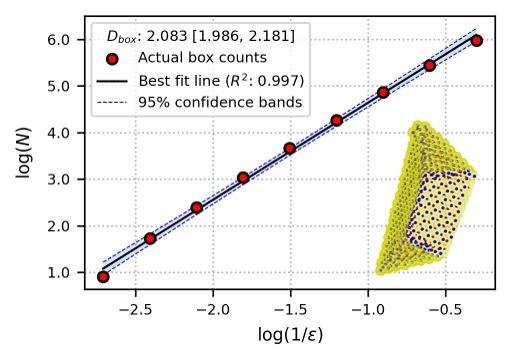} \\
\includegraphics[width=0.45\textwidth]{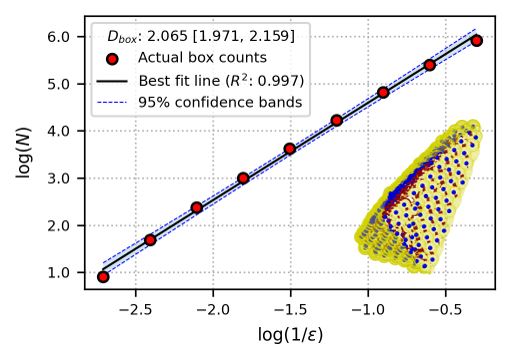}
\includegraphics[width=0.45\textwidth]{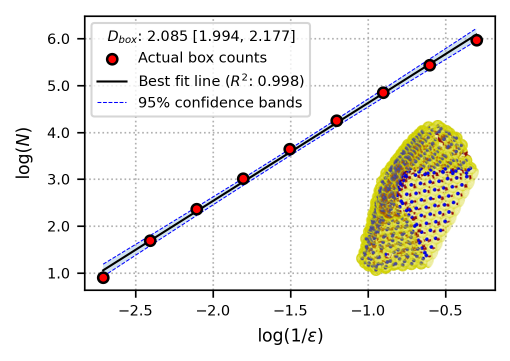} \\
\caption{Full range log-log plots of the linear regression on the box-counting data collected from the voxel representation of tetrahedral palladium nanoparticle surfaces.}
\label{fig:TH_VX_loglogFullRangePlots}
\end{figure}

\begin{figure}[htbp]  \centering
\includegraphics[width=0.45\textwidth]{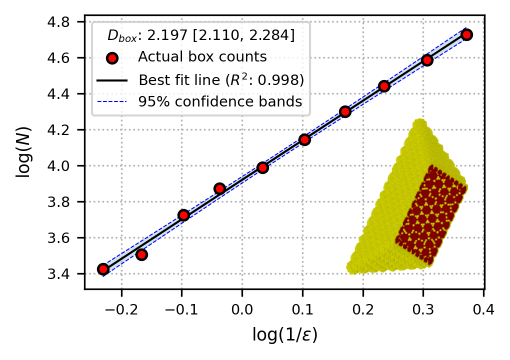}
\includegraphics[width=0.45\textwidth]{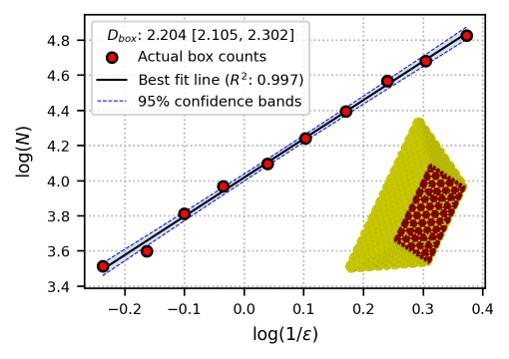} \\
\includegraphics[width=0.45\textwidth]{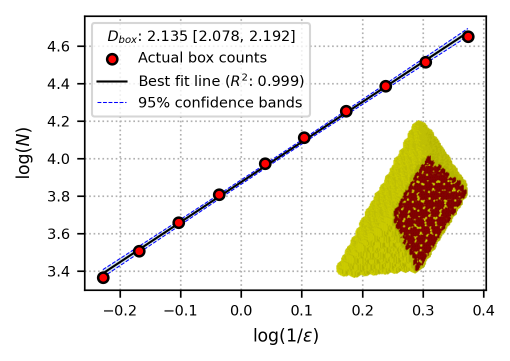}
\includegraphics[width=0.45\textwidth]{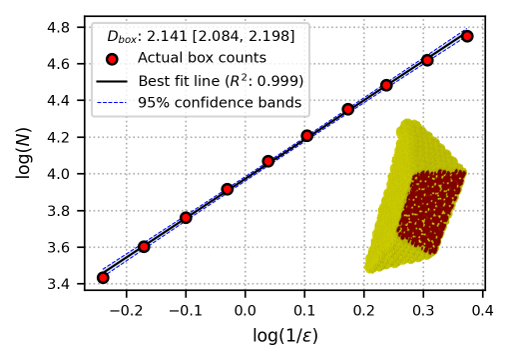} \\
\includegraphics[width=0.45\textwidth]{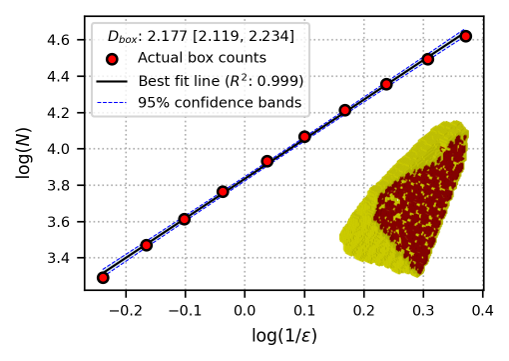}
\includegraphics[width=0.45\textwidth]{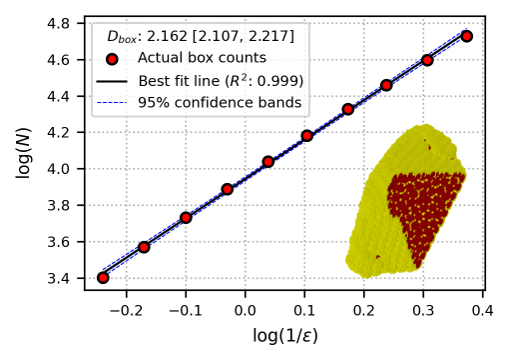} \\
\caption{Full range log-log plots of the linear regression on the box-counting data collected from the exact representation of tetrahedral palladium nanoparticle surfaces.}
\label{fig:TH_EX_loglogFullRangePlots}
\end{figure}

Figures~\ref{fig:OT_VX_loglogFullRangePlots} to \ref{fig:TH_EX_loglogFullRangePlots} show the log-log plots of the box-counting results prior to any point removal. The box-counts for all test cases showed good conformity to linearity.

\begin{figure}[htbp]  \centering
\sidesubfloat[]{\includegraphics[width=0.95\textwidth]{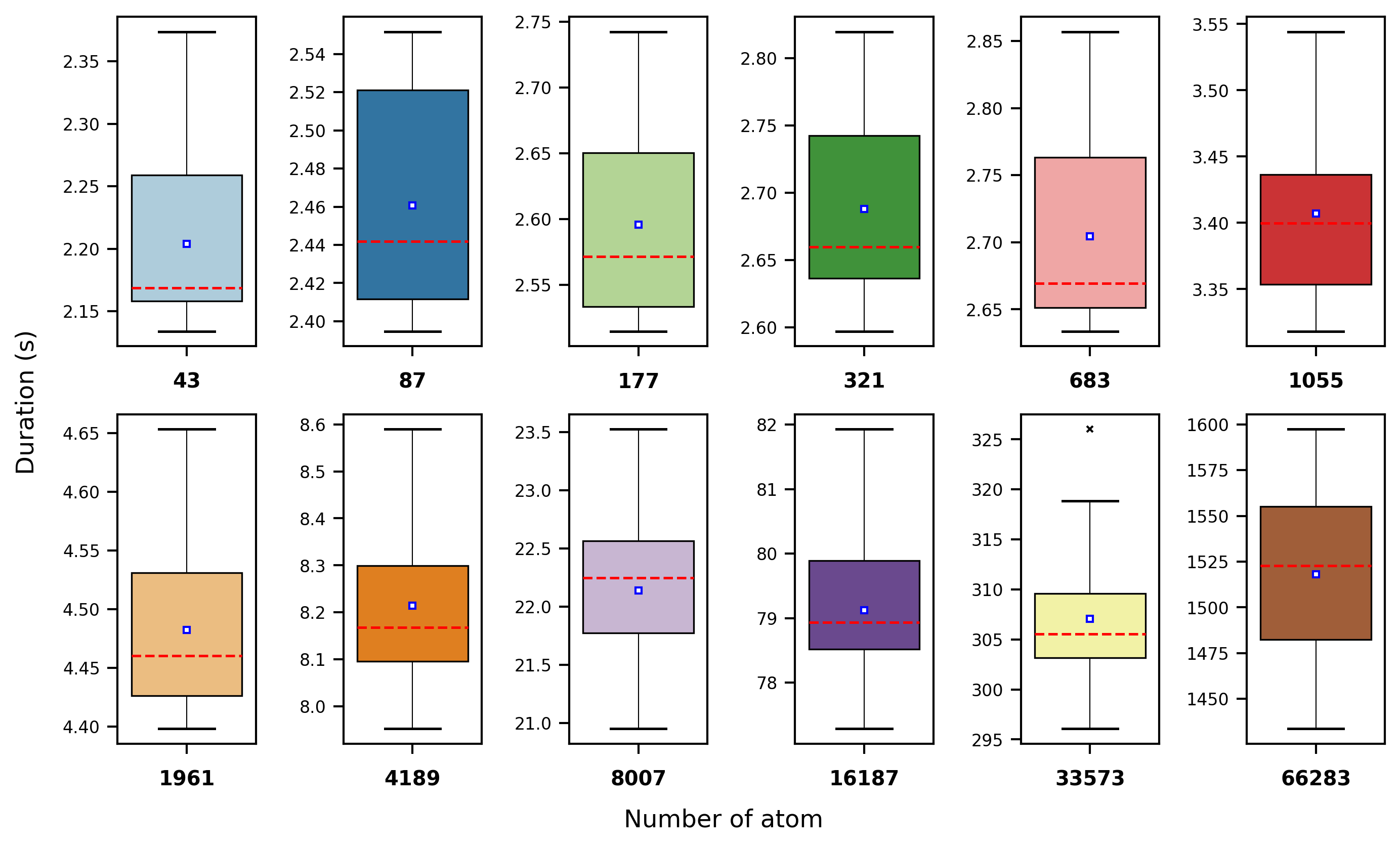}\label{fig:TC_VX_BoxPlots}} \\
\sidesubfloat[]{\includegraphics[width=0.95\textwidth]{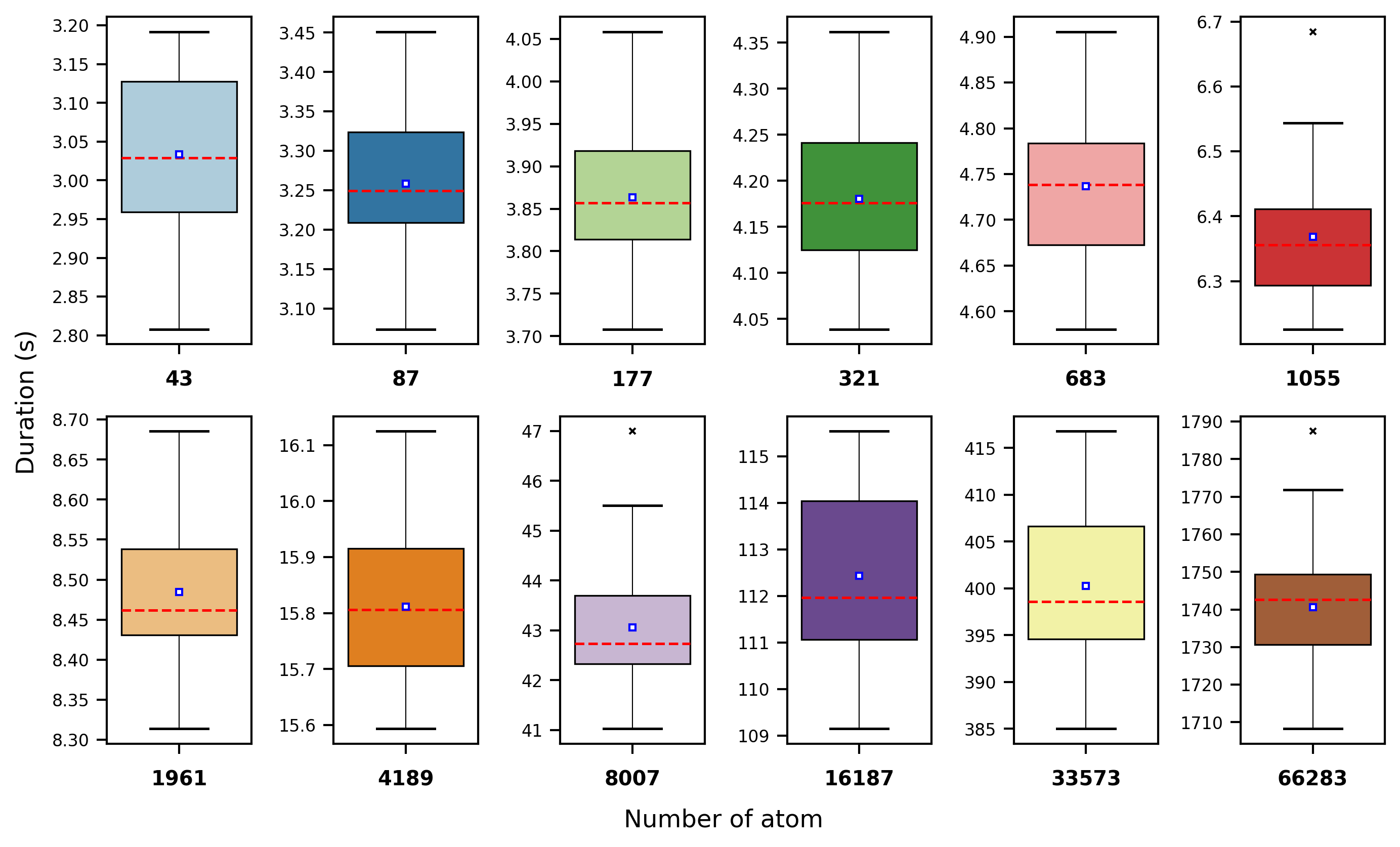}\label{fig:TC_EX_BoxPlots}} \\
\caption{Box plots of algorithm run time data distribution prior to outlier removal (blue dots and red lines represent mean and median, respectively), obtained from the (a-b) time complexity, (c-d) strong scaling, and (e-f) weak scaling tests of the box-counting algorithms for voxelised point cloud (left) and mathematically exact (right) surface representations.}
\end{figure}

\begin{figure}[htbp]  \centering
\ContinuedFloat
\sidesubfloat[]{\includegraphics[width=0.6\textwidth]{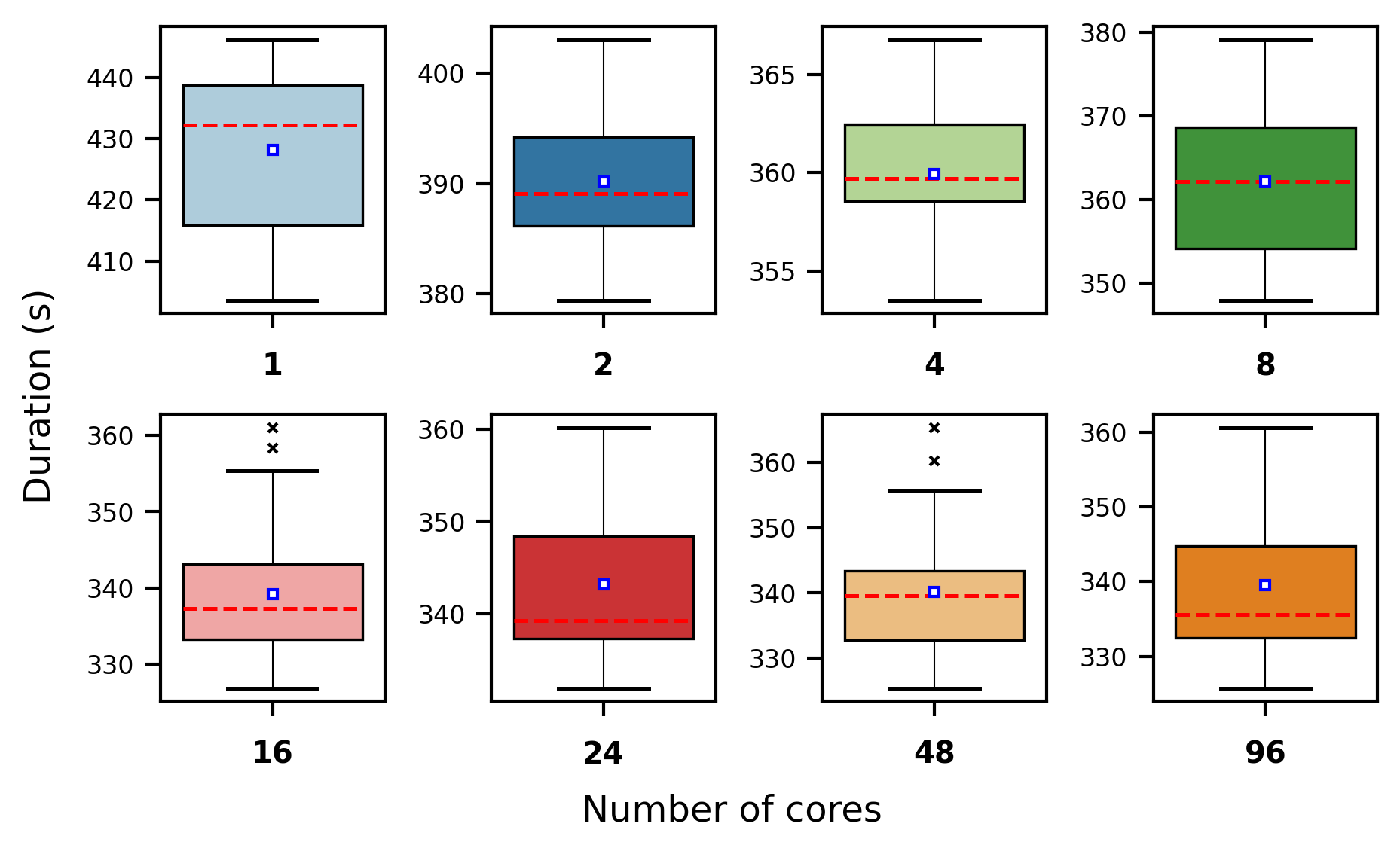}\label{fig:SS_VX_BoxPlots}} \\
\sidesubfloat[]{\includegraphics[width=0.6\textwidth]{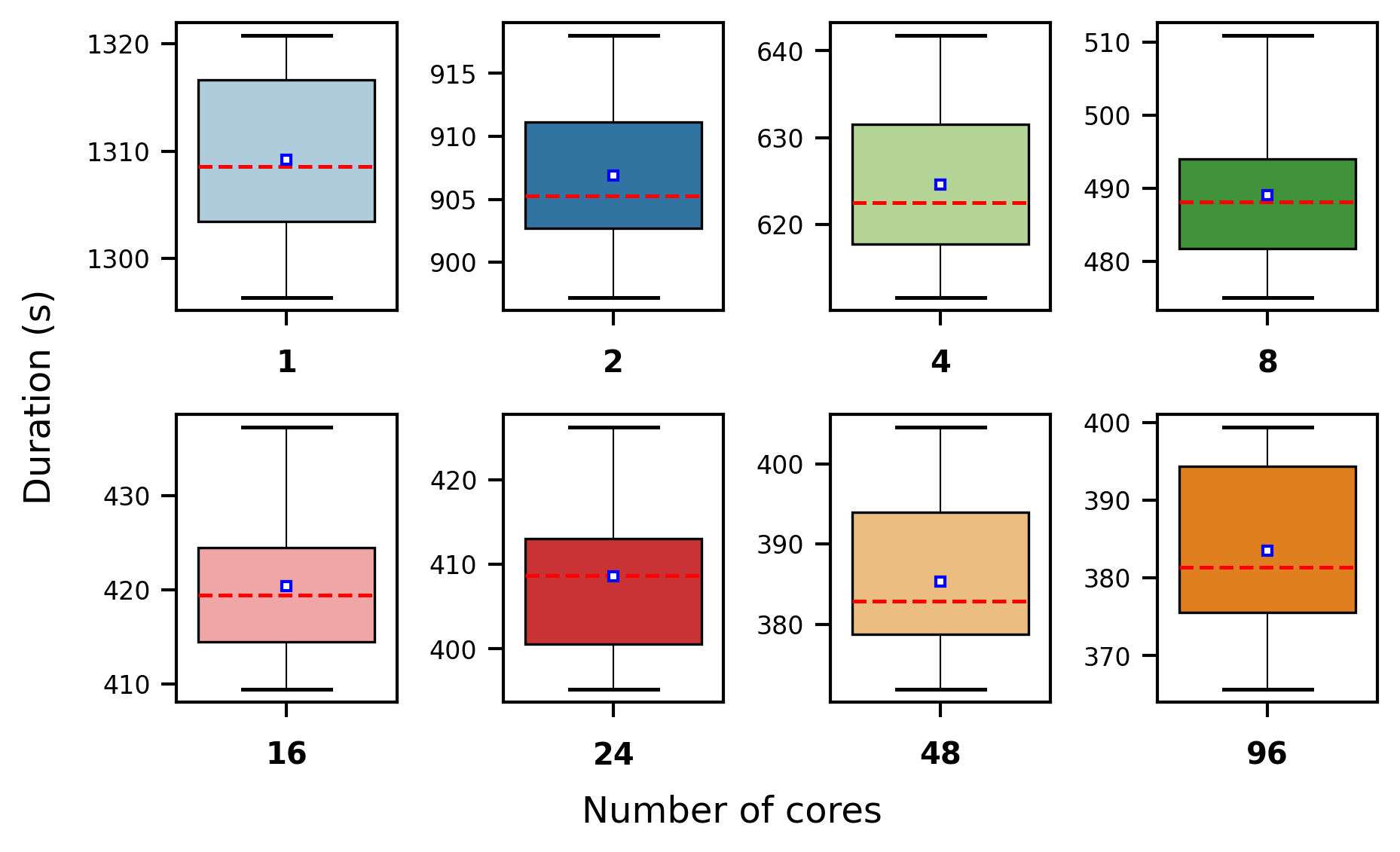}\label{fig:SS_EX_BoxPlots}} \\
\caption{Box plots of algorithm run time data distribution prior to outlier removal (blue dots and red lines represent mean and median, respectively), obtained from the (a-b) time complexity, (c-d) strong scaling, and (e-f) weak scaling tests of the box-counting algorithms for voxelised point cloud (left) and mathematically exact (right) surface representations.}
\end{figure}

\begin{figure}[htbp]  \centering
\ContinuedFloat
\sidesubfloat[]{\includegraphics[width=0.6\textwidth]{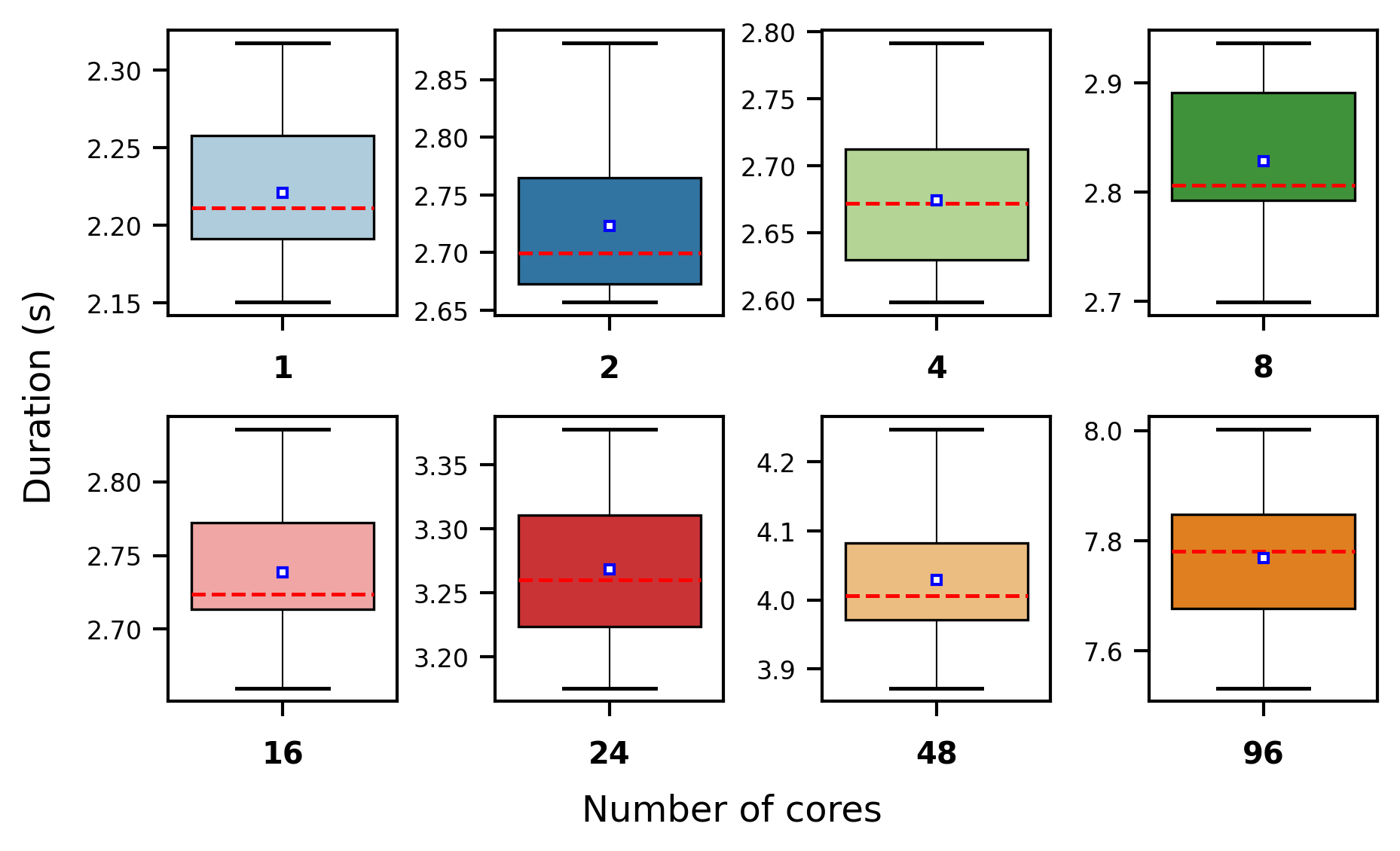}\label{fig:WS_VX_BoxPlots}} \\
\sidesubfloat[]{\includegraphics[width=0.6\textwidth]{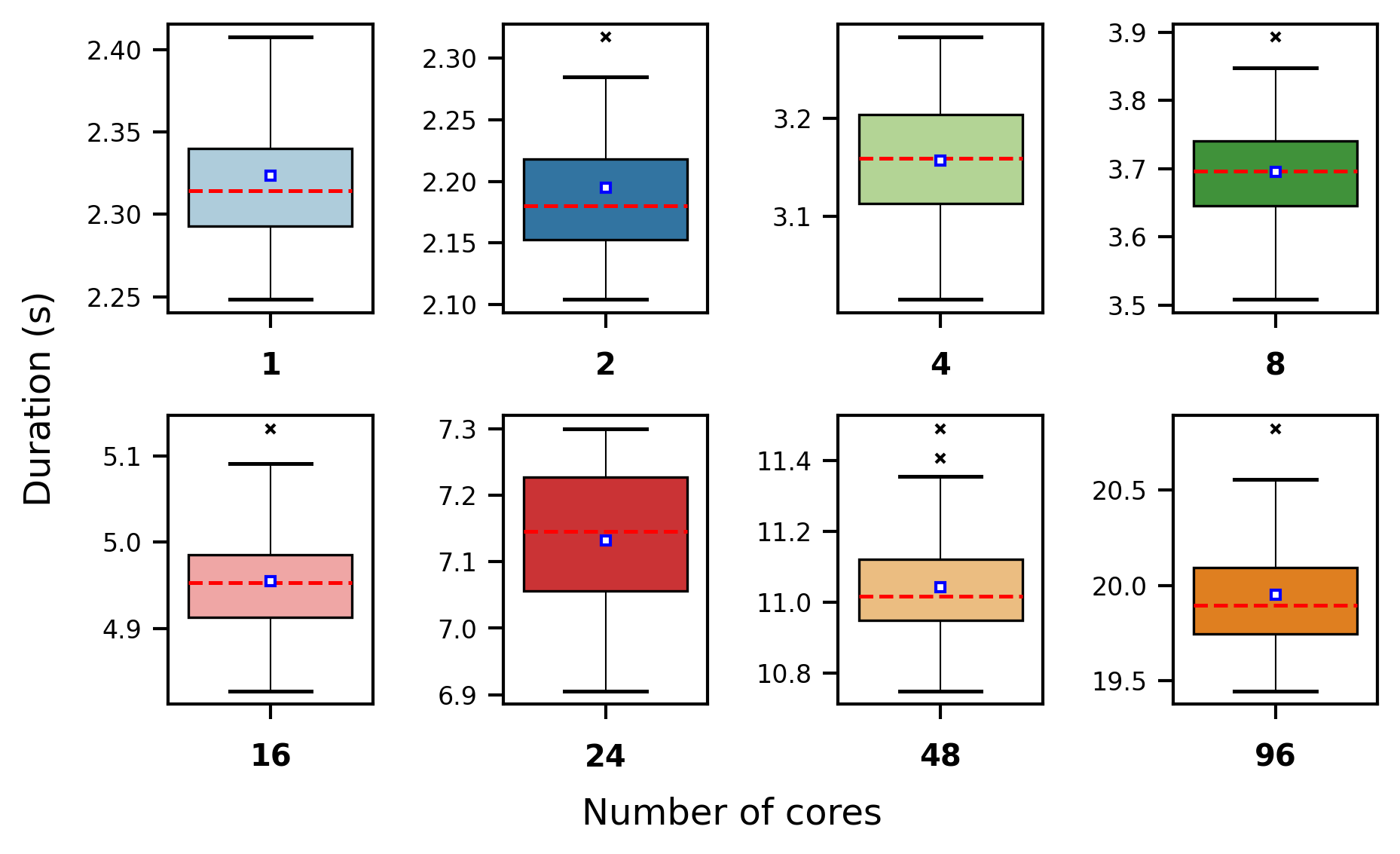}\label{fig:WS_EX_BoxPlots}} \\
\caption{Box plots of algorithm run time data distribution prior to outlier removal (blue dots and red lines represent mean and median, respectively), obtained from the (a-b) time complexity, (c-d) strong scaling, and (e-f) weak scaling tests of the box-counting algorithms for voxelised point cloud (left) and mathematically exact (right) surface representations.}
\label{fig:boxPlots}
\end{figure}

